\documentclass[11pt]{article}
\pdfoutput=1
\usepackage{jheppub,amsmath,amssymb}
\usepackage{enumerate}
\usepackage{bm}
\usepackage{bbm}
\usepackage[usenames,dvipsnames]{xcolor}
\usepackage{graphics}
\usepackage{tikz,todonotes}
\usepackage[utf8]{inputenc}
\usepackage{cancel}
\usetikzlibrary{arrows,positioning,decorations.markings,decorations.pathmorphing,calc}
\usepackage{color}
\usepackage{graphicx}
\usepackage{dcolumn}
\usepackage{bm}
\usepackage{amssymb}
\usepackage{amsmath}
\usepackage{sectsty}
\usepackage{colortbl}
\usepackage{latexsym}
\usepackage{float}
\usepackage{ifthen}
\usepackage{caption,subfig}
\usepackage{enumerate}
\usepackage{url}
\usepackage{caption,subfig}
\usepackage{xfrac}
\usepackage{amsmath}
\usepackage{tensor}
\usepackage{soul}

\definecolor{hyperref}{RGB}{026,028,087}

\newcommand{\mpl}{M_{\rm Pl}}
\renewcommand{\Im}{{\rm Im}}
\renewcommand{\Re}{{\rm Re}}

\def\be{\begin{equation}}
\def\ee{\end{equation}}
\def\ba{\begin{eqnarray}}
\def\ea{\end{eqnarray}}

\def\nn{\nonumber}

\def\d{\mathrm{d}}

\def\ba{\begin{eqnarray}}
\def\ea{\end{eqnarray}}

\def\L{\mathcal{L}}

\def\D{\mathcal{D}}
\def\A{\mathcal{A}}
\def\Z{\mathcal{Z}}

\def\d{\mathrm{d}}
\def\mn{_{\mu \nu}}

\def\({\left(}
\def\){\right)}

\def\mpl{M_{\rm Pl}}
\def\p{\partial}

\begin{document}

\title{Causality in Curved Spacetimes: \\ The Speed of Light \& Gravity}

\author{Claudia de Rham}
\author{\& Andrew J. Tolley}
\affiliation{Theoretical Physics, Blackett Laboratory, Imperial College, London, SW7 2AZ, UK}
\affiliation{CERCA, Department of Physics, Case Western Reserve University, 10900 Euclid Ave, Cleveland, OH 44106, USA}

\emailAdd{c.de-rham@imperial.ac.uk}
\emailAdd{a.tolley@imperial.ac.uk}

\abstract{Within the low-energy effective field theories of QED and gravity, the {\it low-energy} speed of light or that of gravitational waves can typically be mildly superluminal in curved spacetimes. Related to this, small scattering time advances relative to the curved background can emerge from known effective field theory coefficients for photons or gravitons. We clarify why these results are not in contradiction with causality, analyticity or Lorentz invariance,  and highlight various subtleties that arise when dealing with superluminalities and time advances in the gravitational context.  Consistent low-energy effective theories are shown to self--protect by ensuring that any time advance and superluminality calculated within the regime of validity of the effective theory is necessarily {\it unresolvable}, and cannot be argued to lead to a macroscopically larger lightcone. Such considerations are particularly relevant for putting constraints on cosmological and gravitational effective field theories and we provide explicit criteria to be satisfied so as to ensure causality.}

\maketitle


\section{Introduction}

Causality and unitarity play a crucial role in fixing the structure of a Lorentz invariant quantum field theory as was recognized early on. This is most immediately apparent in the dispersion relation methods \cite{Nussenzveig:1972tcd} utilized for example in the spectral representation of K\"allen and Lehmann \cite{Kallen:1952zz,Lehmann:1954xi} where the Fourier space Feynman propagator is recognized to be an analytic function of complex momentum squared up to a pole and right hand branch cut. These dispersion relation methods evolved into the S--matrix analyticity program of the 1960's which --albeit in a different form-- plays a crucial role today in amplitude methods. More recently these ideas have been used to put constraints on low-energy effective theories (EFTs), either through positivity bounds \cite{Pham:1985cr,Ananthanarayan:1994hf,Adams:2006sv,Bellazzini:2016xrt,Cheung:2016wjt,Bonifacio:2016wcb,deRham:2017avq,deRham:2017imi,deRham:2017zjm,Bellazzini:2017fep,deRham:2018qqo,deRham:2017xox,Zhang:2018shp,Melville:2019wyy,Alberte:2019xfh,
Alberte:2019zhd,Kim:2019wjo,Wang:2020jxr,Afkhami-Jeddi:2018own}, demanding scattering time delays are positive (asymptotic (sub)luminality) \cite{Camanho:2014apa,Hinterbichler:2017qyt,Bonifacio:2017nnt,Hinterbichler:2017qcl,AccettulliHuber:2020oou}, or related methods.\\

One unfortunate feature of these methods is that there is no clear way to extend them to curved spacetimes, except perhaps maximally symmetric  cases, see \cite{Baumann:2015nta,Afkhami-Jeddi:2018own,Baumann:2019ghk} for some attempts to deal with this. The powerful analyticity properties combined with crossing symmetry are spoiled when the background spacetime is time-dependent. In the absence of clear causality constraints, one reasonable guess is to demand that relativistic causality should require the propagation speed of all degrees of freedom to be (sub)luminal.  For a non-gravitational theory this is largely a reasonable criterion, and indeed it is known that certain constraints from positivity bounds appear to be connected with (sub)luminality of fluctuations about different backgrounds \cite{Adams:2006sv}. \\

For a gravitational theory these questions are altogether more subtle. The ability to perform field redefinitions that change the off-shell metric means that it is no longer clear which lightcone to use as a reference. Due to the ambiguity of field redefinitions, causality in a gravitational theory is usually phrased for on-shell invariant quantities, such as the requirement that the tree level scattering matrix is analytic in terms of its Mandelstam variables. The relation between causality and analyticity is highlighted for example in Refs.~\cite{Bremermann:1958zz,bogoliubov1959introduction,Hepp_1964}. But such considerations are typically of little use in understanding causality in spacetimes with significant curvature effects such as FLRW and Schwarzschild.
The peculiar subtleties associated with causality in curved spacetime are well known in the context of the low-energy effective theory for QED below the electron mass \cite{Drummond:1979pp,Lafrance:1994in}, where the low-energy phase and group velocity of light is known to be superluminal for certain polarization states on some curved geometries. The relation with analyticity and causality in this example has been discussed extensively in the literature, see for example Refs.~\cite{Hollowood:2007kt,Hollowood:2007ku,Hollowood:2008kq,Hollowood:2009qz,Hollowood:2010bd,Hollowood:2010xh,Hollowood:2011yh,Hollowood:2012as,Hollowood:2015elj,Goon:2016une}. \\

From the cosmological perspective, it is known that various cosmological models such as inflation and dark energy theories can easily exhibit different speeds of propagation in different sectors. It is also known that the speed of gravitational waves (GWs) in a given low-energy EFT can be different than the luminal speed inferred from the metric from which the theory is constructed. Once again this leads to the obvious question of how causality should be set in the cosmological context. In the literature, it is often considered, by fiat, that the speed of GWs and of all other species should be (sub)luminal in an arbitrary background with respect to the metric out of which the theory is constructed. Doing this often imposes  constraints on the signs of coefficients in the effective action, but as already mentioned such a criterion is not invariant under field redefinitions. \\

The previous criterion can sometimes be correct, however in the canonical case of gravitational effective theories it is not necessarily the case, and indeed {\bf{demanding $c_s \le 1$ for {\it{all}} species can sit in contradiction with those same requirements of causality and analyticity}}. In two recent papers \cite{deRham:2019ctd,deRham:2020ejn} it was pointed out that integrating out matter fields typically leads to curvature operators in the low-energy EFT of gravity that can lead to a (small!) superluminal low-energy speed for GWs. For the low-energy EFT of gravity, constraining the signs of the low-energy operators to be so as to entirely forbid superluminal low-energy speed, no matter how small, would lead to a criterion in contradiction with known partial high energy completions and more generally expectations based on analyticity. Instead, we highlight that a small amount of low-energy superluminal speed does not directly imply that the support of the retarded propagator lies outside the usual lightcone.
 \\

The key point is that superluminality of the low-energy speed, is only in conflict with relativistic causality if it can be integrated over time to make a large macroscopic effect, i.e. the light cone of causal influence is {\bf measurably larger} than the background geometry lightcone. In solving for the retarded propagator perturbatively in the EFT expansion around the background one, this only occurs when there are {\bf secular} effects that need to be resummed, leading to significant differences in the structure of the propagator at late times (or large distances).
We will demonstrate that in those situations where a small superluminal low-energy speed arises in a given gravitational effective theory, the requirement that the EFT is under control automatically precludes any secular effects, preserving causality by ensuring that at any finite order in the convergent EFT expansion, the retarded propagator has the same causal structure as the unperturbed one. That this is the case hinges on the smallness of the superluminal speed correction that arises within gravitational effective theories. We may regard this as a `self-protection' mechanism against causality violation for consistent low-energy EFTs.
\\

In the case of asymptotically flat geometries where an S-matrix may be defined this discussion may be further sharpened. We may precisely define a relativistic generalization of the Eisenbud-Wigner scattering time-delay $\Delta T$, \cite{Eisenbud:1948paa,Wigner:1955zz,Smith:1960zza,Martin:1976iw}. In particular for spherically symmetric spacetimes we may define a time delay $\Delta T_\ell$  for each partial wave $\ell$ via the derivative of the scattering phase shift at fixed $\ell$
\be
\Delta T_\ell = 2\frac{\partial  \delta_\ell(\omega) }{\partial \omega}\Big|_{\ell}\, ,
\ee
in terms of the incident particle's energy $\omega$. We may also consider the inequivalent but related time delay for fixed impact parameter $b$
\be
\Delta T_b = 2\frac{\partial  \delta_\ell(\omega) }{\partial \omega}\Big|_{b}\, .
\ee
In General Relativity (GR) these time-delays are the well--known `Shapiro' or `gravitational' time-delays \cite{Shapiro:1964uw}. There is a long history of using positivity or boundedness\footnote{In non-relativistic quantum mechanics, the time delay can be negative, as in the case of hard sphere scattering. The original requirement of Wigner is only that the magnitude of the associated time advance is bounded \cite{Wigner:1955zz}.} of this time delay as a means of imposing causality (see for example \cite{DECARVALHO200283} for a review) going back to Eisenbud and Wigner and improved by Smith \cite{Smith:1960zza}. More recently in the context of relativistic field theories, in \cite{Camanho:2014apa} the eikonal approximation was used to determine this scattering time-delay in various low-energy effective theories. Subsequently, for example in \cite{Camanho:2016opx,Hinterbichler:2017qyt,Bonifacio:2017nnt,Hinterbichler:2017qcl,DAppollonio:2015fly,AccettulliHuber:2020oou},
it was generally argued on the same lines outlined by Eisenbud and Wigner \cite{Eisenbud:1948paa,Wigner:1955zz} that an overall negative time-delay is a signal of causality violation and inconsistency of the low-energy effective theory. This criterion is sometimes dubbed absence of `asymptotic superluminality' \cite{Gao:2000ga}.
The perspective of some literature  is that any time delay for which there is net negative sign $\Delta T<0$, i.e. a net time advance, leads to a causality violation, since it implies propagation faster than the asymptotic spacetime, regardless of magnitude, \cite{Camanho:2016opx}. \\

In the case of weakly coupled UV completions, it was argued in Ref.~\cite{Camanho:2014apa} that apparent time-advances in the low-energy effective theory could be resolved by an infinite tower of higher spins as in the case of string theory. While useful for understanding the nature of possible UV completions, from the perspective of a low-energy physicist this tower of higher spins would just show up as an infinite number of local operators.  Clearly including more and more irrelevant operators in the low-energy EFT cannot change the statement that the phase shift (as computed using the leading order operators in the EFT) is of a particular sign at sufficiently low-energy.  Furthermore since the low-energy effective theory is designed to describe large distance physics, it should reliably compute the large distance behaviour of the retarded propagator. Any time advance in addition to the Shapiro contribution would signal that the retarded propagator has support outside the lightcone set by the background geometry. Whether or not the UV completion admits an infinite tower of spins cannot by itself resolve this tension with causality at large distances  which is where the real issue lies. \\

Hence the import of the observation that the UV theory resolves the time advance for the low-energy physicist is merely the generic statement that the low-energy effective theory has a cutoff (set by the lowest mass in the tower for a weakly coupled UV completion), and only a {\bf resolvable} time advance calculable within the regime of validity of the effective theory would signal a true causality violation. This criterion is of course true regardless of whether or not we consider a weakly coupled UV completion. This implies that, from what the low-energy EFT is concerned, the resolution behind a negative sign of the phase shift in an EFT and its apparent tension with causality cannot lie in the existence of a higher-spin tower per se but rather must lie in the actual order of magnitude of the phase shift/time advance itself and the existence of a cutoff irrespectively of what precisely happens at that cutoff (irrespectively on whether or not it represents the onset of a tower of higher spins or whether it represents instead the mass of other heavy particles whose loops are relevant.)\\

Moreover, the causality criterion proposed in \cite{Camanho:2014apa} is only that the {\bf total} time delay $\Delta T$ is positive within the regime of validity of the effective theory $\Delta T>0$. This is the essential point of the `asymptotic superluminality' condition \cite{Gao:2000ga}, local perturbations and corrections to the effective theory may speed up local propagation relative to the background geometry, but this is viewed as acceptable as long as they remain slower than the asymptotic geometry. \\

 Our perspective is that this is incorrect, or at least incomplete, for two reasons (a) strict positivity of the total time delay fails to account whether the delay is resolvable, and (b) more generally what is required is (resolvable) positivity of the EFT corrections to the time delay {\bf relative to the background}\footnote{At least in cases in which there is a clean $\mpl \rightarrow \infty$ decoupling limit, and for which the Shapiro contribution is defined in the field frame for which the decoupling limit is well defined, to be discussed in section~\ref{sec:DL}.}. This may be seen by a more careful consideration of the well known example of QED in curved spacetime. It is well known that loop corrections from charged particles induce low-energy superluminal propagation for photons in a curved background, e.g. for a Schwarzschild background, as first noted by Drummond and Hathrell \cite{Drummond:1979pp}. The total time delay is then
\be
\Delta T = \Delta T^{g} + \Delta T^{\rm EFT} \, ,
\ee
where $ \Delta T^{g} $ is the time-delay induced by the curved background spacetime, i.e. the Shapiro time-delay, and $\Delta T^{\rm EFT}$ is in this case the Drummond-Hathrell correction from loop effects.
According to \cite{Camanho:2014apa} this is consistent with causality because it is noted that in order for the negative $\Delta T^{\rm EFT}$ to overpower the positive (in dimensions $D>4$) $\Delta T^{g}$, it is necessary that the impact parameter is smaller than the inverse mass of the charged particle integrated out, which means the low-energy effective theory can no longer be trusted. However, this argument does not resolve the causality problem that appears to arise at larger impact parameter where $\Delta T^{\rm EFT}$ is negative, and its calculation can be trusted within the regime of validity of the EFT,  but the overall $\Delta T$ is positive. Any negative $\Delta T^{\rm EFT} $ implies that the photons are travelling superluminally relative to the background metric (accounted for by the Shapiro contribution) and this sits in clear contradiction with the fact that in the known UV completion, namely QED in curved spacetime, causality remains intact with the causal lightcone defined by the background geometry and not the asymptotic Minkowski geometry. Causality in this case should then be a statement about $ \Delta T^{\rm EFT}$ itself, and not about the full $\Delta T$. Yet demanding  $ \Delta T^{\rm EFT}>0$ sits  in contradiction with the well-known result of Drummond and Hathrell. \\

Hence the `asymptotic' causality condition of \cite{Camanho:2014apa} fails to address what is the resolution of the apparent causality violation in the low-energy Drummond-Hathrell EFT. This case was discussed in details in Ref.~\cite{Goon:2016une}. As we shall demonstrate later, and as was cleanly argued in \cite{Hollowood:2015elj} for calculations in the shockwave (eikonal) limit, the real resolution of causality is that the negative Drummond-Hathrell contribution is {\bf unresolvable}\footnote{Some literature use the cutoff to define resolvability, this is not the definition we will use in what follows for reasons explained around \eqref{eq:naivebound}.} within the effective theory, namely within the regime of validity of the EFT it remains always true that the advance is smaller than the {\bf `geometric optics' resolution scale}
\be
|\Delta T^{\rm EFT} | \ll \omega^{-1}\,,
\ee
where $\omega$ is the asymptotic energy of the scattering particle.  Equivalently this is the statement that the EFT contribution to the scattering phase shift is bounded by unity
\be
  |\Delta \delta_\ell^{\rm EFT}| \ll 1 \, .
\ee
It is therefore clear that by itself the sign of the time-delay correction cannot be sufficient in determining whether or not acausality will follow, it is crucial to consider also its magnitude. Only if there is a time advance, calculable within the EFT, larger than the wavelength of the scattering state, can we infer genuine causality violation.  The eikonal approximation used in \cite{Camanho:2014apa}\footnote{In Ref.~\cite{Camanho:2014apa} as well as in for example the nice recent related discussion of \cite{AccettulliHuber:2020oou} derivations are principally performed within the eikonal approximation which has the virtue of having a relatively clean interpretation in terms of Feynman diagrams as a resummation of ladder diagrams. As such, these derivations are well suited to perturbative S-matrix calculations. All of the calculations we perform here are in the related semiclassical approximation. This approximation is harder to understand in terms of a resummation of Feynman diagrams, however it is straightforward to calculate it by applying the WKB approximation to the corrected Green's functions. This is the method we will use in what follows and we refer to Appendix~\ref{app:phaseshift} for more details on the semiclassical approximation. The eikonal approximation may be obtained from a high energy limit of the semiclassical approximation, as we outline in Appendix~\ref{app:eikonal}. In the relativistic context the latter may be viewed as a Penrose limit of the former. Due to the close connection many of our statements apply implicitly to both the semiclassical and eikonal approximation methods.}  relies on resumming an infinite number of ladder diagrams which are ``typically" enhanced as compared to all the other diagrams. Implicitly this resummation amounts to an exponentiation of the lowest order phase shift in the partial wave expansion
\be
\delta_\ell =\delta^g_\ell+ \Delta \delta_\ell^{\rm EFT} \rightarrow \frac{1}{2i} \left( e^{2i (\delta^g_\ell+ \Delta \delta_\ell^{\rm EFT} )}-1 \right) \, .
\ee
More precisely, it is the statement that a resummation of the $t-$channel exchange and its associated higher order ladder diagrams appropriately exponentiate in the manner
\be
\delta_\ell(\text{$t-$channel}) + \sum \text{ladder diagrams}=  \frac{1}{2i} \left( e^{2i \delta_\ell(\text{$t-$channel})}-1 \right)\,.
\ee
This resummation does make sense when the enhancement does actually occur, or in other words when the resulting phase-shift is $|\delta_\ell|\gtrsim 1$ so that these terms may be taken large relative to other small corrections. However, as we shall demonstrate, the validity of the low-energy effective theory actually requires $| \Delta \delta_\ell^{\rm EFT}|  \ll 1$ in the case of QED  and we should really replace the exponentiated form with
\ba
e^{2i (\delta^g_\ell+ \Delta \delta_\ell^{\rm EFT} )} \rightarrow e^{2i \delta^g_\ell } \left\{ 1+ 2 i \Delta \delta_\ell^{\rm EFT} + \( \frac{1}{2} (2i)^2 (\Delta \delta_\ell^{\rm EFT})^2 +
\begin{array}{c}\text{terms of the same order} \\
\text{ in EFT expansion}
\end{array}
\)\right\} \,. \nn
\ea
Although the eikonal resummation is valid for the usual $\delta^g_\ell$ contributions, we cannot take seriously the exponentiated $\Delta \delta_\ell^{\rm EFT} $ contributions as an indicator of high energy behaviour which is necessary in order to interpret them as contributing to a physical time delay. Explicit calculations in the UV theory such as those performed in \cite{Hollowood:2015elj} confirm that eikonal resummed Drummond--Hathrell result bears no relation to the true high energy behaviour of the scattering phase shift. \\

We shall argue that this is a more general phenomena, applying equally for GWs and indeed to any EFT. Naive scattering time advances do occur in the low-energy effective theory of gravity, and generic EFTs in curved spacetime, arising for example in Schwarzschild spacetime from matter loops in close analogy with the QED case. However as we show later these time advances are not resolvable and are seen to satisfy
\be
|\Delta T^{\rm EFT} | \ll \omega^{-1} \, ,
\ee
which is equivalent to the statement that
\be
|\Delta \delta _\ell^{\rm EFT} | \ll 1 \, .
\ee
This shows up generically in the fact that local fluctuations about the background geometry allow for mildly superluminal fluctuations. Once again though, this superluminality is not resolvable, and cannot be used to argue for any causality violations. By contrast, genuinely acausal EFTs lead to resolvable time advances calculable within the regime of validity of the EFT and we give an example later.
Consistent EFTs are seen then to self-protect from any macroscopic causality violation, and in this sense contain remnant information of their consistent UV completion. We stress again that the precise nature of the UV completion is immaterial to this particular part of the argument. The previous bounds and `self-protection' mechanism should indeed hold irrespectively of whether we are dealing with a  weakly coupled infinite tower of spins, or a more mundane heavy loop contribution as in the Drummond--Hathrell case.
\\

These observations have significant impact on how we put constraints on low-energy effective theories. The overly enthusiastic low-energy physicist who demands that the Wilson action should be constrained by the requirement that all fluctuations around every background should be (sub)luminal relative to the background, or similarly that the scattering time delay correction relative to the background (Shapiro/gravitational) time delay is positive may easily risk ruling out EFTs with consistent Lorentz invariant and causal UV completions. A more nuanced discussion is required that establishes whether either of these effects lead to a macroscopically observable causality violation within the regime of validity of the effective field theory. In what follows we give well--known examples of both situations. Related discussions in the holographic context were also considered in \cite{Caceres:2019pok}.\\

The rest of the manuscript is organized as follows:
In section~\ref{sec:refractiveindex} we review why causality and analyticity typically require a subluminal low-energy phase velocity, while pointing out some caveats that occur in curved spacetimes.
In section~\ref{sec:decouplingLim}, we highlight subtleties that arise when dealing with superluminal low-energy speed in gravitational EFTs and the relevance of being able to take a decoupling limit where the gravitational degrees of freedom decouple before being able to restrict low-energy coefficients based on superluminal criteria. We highlight cases where macroscopic superluminalities are allowed (and even sometimes imposed) by analyticity and causality. Such cases are particularly important when attempting to restrict the allowed coefficients in cosmological EFTs.
 In section~\ref{sec:cosmology} we show how the small amount of superluminal low-energy speed we expect from the EFT of gravity leads to no physical propagation outside the light-cone and is therefore not in contradiction with causality. The same type of arguments is then shown to apply for Black Hole (BH) spacetimes in section~\ref{sec:BH}.
 These BH spacetimes are asymptotically flat and the connection with the sign and magnitude of the scattering phase shift can be made manifest within the EFT of gravity. The same type of arguments and absence of secular growth is also made explicit in the EFT of QED below the electron mass as highlighted in section~\ref{sec:QED} where we make it clear that a negative phase shift of sufficiently large magnitude to be in tension with causality can never be realized within the regime of validity of the low-energy QED EFT. This is contrary to what occurs in other EFTs where the semiclassical or eikonal approximation can remain under control for sufficiently large phase shift and hence lead to a resolvable physical time advance and be in tension with causality as illustrated in section~\ref{sec:PX_resolvable}.
 We end with a summary and discussion in section~\ref{sec:discussion}. Appendix~\ref{app:EFT}
  provides a review of the low-energy EFT for gravity
 as well as the graviton dispersion relation and the direction of the RG flow. The relations between the semiclassical approximation used to computing the phase shift and time delay are reviewed in Appendix~\ref{app:phaseshift} and the relation to the eikonal approximation is outlined in Appendix~\ref{app:eikonal}. Finally Appendix~\ref{app:Timedelaycalc} provides useful formula to compute the EFT corrections to the time delay relative to generic effective backgrounds.

\section{Causality and Analyticity}\label{sec:refractiveindex}

\subsection{Refractive Index}\label{sec:refractive}
\paragraph{Implications from Analyticity and Unitarity:} It is a familiar result that the speed of propagation of a wave in a medium is in general different than in vacuum. For instance for a rotational and translation invariant medium, it is sufficient to describe the propagation speed through the refractive index $n(\omega)$ for which the phase velocity is given by $v_p = c/n(\omega)$. The speed of propagation of a wavefront is determined by the front velocity $v_f$ which is given by (from now on we set $c=1$),
\be
v_f = \lim_{\omega \rightarrow \infty} 1/n(\omega) \, .
\ee
Relativistic causality demands that $v_f \le 1$. However this does not preclude the possibility that the low-energy phase or group velocity is superluminal. Superluminal group velocities in particular are a well studied experimental phenomena \cite{Milonni,Brillouin,PhysRevLett.93.203902,deRham:2014zqa} and do not in any way contradict causality, they rather indicate the failure of group velocity as a useful concept in a dispersive medium. This was recognized long ago by Sommerfeld and Brillouin \cite{Brillouin} who resolved the apparent paradox between superluminal group velocities and relativity well before any experimental evidence for this phenomena. \\

However the front velocity is not the end of the story. The full requirement of causality is that the retarded propagator vanishes outside of the forward lightcone,
\be
G_{\rm ret}(x,x') =0 \, , \quad \text{ for }\quad (x-x')^2>0, \quad \text{ or } \quad (t-t') <0\, .
\ee
In addition to the front velocity being luminal, the latter generally requires that the refractive index is an analytic function in the upper half complex $\omega$ plane \cite{Toll:1956cya}. Applying Cauchy's theorem assuming {\bf analyticity} leads to the Kramers--Kronig relations which for future comparison is most usefully written as
\be\label{KKdispersion}
n(\omega)  = n(\infty) + \frac{2}{\pi} \int_0^{\infty} \d \omega' \frac{\omega' {\Im[n(\omega')]}}{(\omega')^2- (\omega+i \epsilon)^2}\,.
\ee
A travelling wave moving in the $z$ direction takes the form $e^{i \omega t- i n(\omega) \omega z}$ and so it is the real part $\Re[n(\omega)]$ that determines the speed, and the imaginary part the dispersion.
Now in a normal medium, {\bf unitarity} demands that the imaginary part of the refractive index is positive: $ \Im[n(\omega)] \ge 0$ for real $\omega>0$. At zero frequency,
the real part  is given more precisely as
\be
\Re[n(0)] =   n(\infty)+ P\left[\frac{2}{\pi} \int_0^{\infty} \d \omega' \frac{ \omega' {\Im[n(\omega')]}}{(\omega')^2}\right]>n(\infty) \, .
\ee
Hence we conclude a bound on the low-energy phase velocity
\be\label{phasefront}
v_p(0)<v_f \, .
\ee
Since the front velocity cannot be superluminal $v_f \le 1$, it is typically inferred that the low-energy phase velocity cannot be superluminal unless we violate either (1) analyticity, or (2) unitarity. \\

\paragraph{Low-energy effective theory:} This particular argument is strengthened if we imagine a situation in which the dispersive imaginary part of the refractive index is only non-zero for frequencies above some scale $M$, i.e.
\be
\Im[n(\omega)] \approx 0 \,,  \quad 0 \le \omega < M \, .
\ee
In this situation, there exists a low-energy effective theory valid for frequencies $\omega \ll M$ for which the refractive index is well approximated by a Taylor series
\be\label{Taylor1}
n(\omega)  = n(0)+ \sum_{n=1}^{\infty}b_n \frac{ \omega^{2n}}{M^{2n}} \, ,
\ee
with leading low-energy term
\be
n(0)=n(\infty) + \frac{2}{\pi} \int_M^{\infty} \d \omega' \frac{\omega' {\Im[n(\omega')]}}{(\omega')^2} > n(\infty) \, ,
\ee
and positive dimensionless EFT coefficients
\be
b_n = M^{2n} \frac{2}{\pi} \int_M^{\infty} \d \omega' \frac{\omega' {\Im[n(\omega')]}}{(\omega')^{2n+2}} >0 \, .
\ee
The equation of motion that describes the propagation of the wave with amplitude $\phi(t,\vec x)$ is of the form
\be\label{LEwaveequation}
n(0)^2 \frac{\partial^2 \phi }{\partial t^2}- \nabla^2\phi = \sum_{n=1}^{\infty }\frac{c_n}{M^{2n}} \frac{\partial^{2n+2} \phi }{\partial t^{2n+2}} \, ,
\ee
for some dimensionless coefficients $c_n$.
The higher time-derivatives arise here due to the low-energy expansion and do not imply additional states. Indeed within the context of the $1/M^2$ expansion this equation may be rewritten in the conceptually nicer form
\be\label{LEwaveequation2}
\frac{\partial^2 \phi }{\partial t^2}- \frac{1}{n(0)^2} \nabla^2\phi = \sum_{n=1}^{\infty }\frac{\tilde c_n}{M^{2n}} \nabla^{2n+2}\phi \, .
\ee
In this low-energy regime, there is by assumption no dispersion, and effects from high energy physics are captured by local higher derivative operators. The leading order group velocity is the same as the phase velocity $v_g =1/n(0)$. Thus low-energy sources propagate at the speed $1/n(0)$. If we compute the retarded propagator for $\phi$ as a perturbative expansion in $1/M$ then each term at finite order will vanish only outside the forward lightcone defined by
\be
(t-t')^2 -n(0)^2 (\vec x-\vec x{\,}')^2 >0 \, ,  \quad t-t'>0 \, .
\ee
Thus we must have $n(0) \ge 1$, otherwise even a tiny superluminal velocity $n(0)= (1 + \epsilon)^{-1}$  with $\epsilon>0$ would integrate up over sufficiently long periods of time to an arbitrary large increase $\Delta x$  in the spatial size of the causal lightcone from a given event $\Delta x = \epsilon |t-t'|$, leading to causality paradoxes. In running this argument, it is crucial that $|t-t'|$ may be made arbitrarily large. We will see that when considering the same argument in a Friedmann--Lema\^itre--Robertson--Walker (FLRW)
geometry or on the background of a BH, it is exactly this assumption that breaks for reasons to be explained. Furthermore it is unclear whether \eqref{phasefront} holds in curved spacetimes due to the generic absence of conventional analyticity.\\

The fact that the situation is more subtle in curved spacetimes is well known from work on the low-energy effective theory for QED in curved spacetime \cite{Drummond:1979pp,Lafrance:1994in} which has been discussed extensively in the literature \cite{Hollowood:2007kt,Hollowood:2007ku,Hollowood:2008kq,Hollowood:2009qz,Hollowood:2010bd,Hollowood:2010xh,Hollowood:2011yh,Hollowood:2012as}, where it is noted that the low-energy phase velocity in a curved spacetime can be superluminal without contradicting the requirement that the front velocity is luminal. Our subsequent discussions in sections~\ref{sec:cosmology}, \ref{sec:BH} and Appendix~\ref{app:EFT} will parallel this for the speed of GWs themselves.

\subsection{Analyticity with Gravity}

Since one of our principal interests is the speed of gravity, i.e. the speed of GWs in a curved background, we would ideally repeat the argument of the previous section. A knowledge of the spectral properties of the propagator for GWs in a curved spacetime could allow us to infer concrete statements about the low-energy speed. Unfortunately a direct application of these arguments to curved spacetimes is not available since there is no requirement that analyticity should hold in general. For cosmological spacetimes, this is made transparent by the inherent time dependence of the background meaning that frequency $\omega$ is no longer a good Fourier variable. \\

Fortunately all information from analyticity in Minkowski spacetime is not completely lost. Consider a particular diffeomorphism invariant low-energy gravitational theory. Analyticity constraints will impose restrictions on the form of low-energy action based on analytic scattering amplitudes in Minkowski spacetime or spectral density requirements. Since the underlying gravitational theory is diffeomorphism invariant, this can immediately be used to infer constraints on covariant operators in the effective Lagrangian which in turn have consequences around curved spacetimes. One set of arguments of this kind are reviewed in appendix \ref{app:EFT} for which we summarise the essential points here. These arguments are closely related to positivity bound arguments that apply to scattering amplitudes \cite{Pham:1985cr,Ananthanarayan:1994hf,Adams:2006sv,Bellazzini:2016xrt,Cheung:2016wjt,Bonifacio:2016wcb,deRham:2017avq,deRham:2017imi,deRham:2017zjm,Bellazzini:2017fep,deRham:2018qqo,deRham:2017xox,Zhang:2018shp,Melville:2019wyy,Alberte:2019xfh,Alberte:2019zhd,Kim:2019wjo,Wang:2020jxr}. \\

Rather than working with a dispersion relation for the refractive index \eqref{KKdispersion}, we can determine a K\"allen-Lehmann type dispersion relation for the gravitational wave propagator on Minkowski spacetime. Due to gauge invariance, this is most conveniently expressed as the exchange interaction `$TT$-amplitude' between two conserved sources with gravitational propagator as given in \eqref{propagator2},
\ba \label{TTamp}
\Delta S_{TT} =\int \frac{\d^4 k}{(2 \pi)^4} &\Bigg[ &\frac{Z}{2 \mpl^2}\frac{(|T_{\mu\nu}(k)|^2-\frac{1}{2} |T(k)|^2)}{k^2-i \epsilon} \\
&+&\frac{C_2(\mu_0)}{\mpl^4} (|T_{\mu\nu}(k)|^2-\frac{1}{3} |T(k)|^2)
+ \frac{C_0(\mu_0)}{\mpl^4} |T(k)|^2 \nn\\
&+&\frac{1}{\mpl^4}  \int_0^{\infty} \d \mu \rho_2(\mu) \frac{(\mu-k^2)}{(\mu+\mu_0)}\frac{(|T_{\mu\nu}(k)|^2-\frac{1}{3} |T(k)|^2)}{\mu+k^2-i \epsilon} \nn  \\
&+&  \frac{1}{\mpl^4}\int_0^{\infty} \d \mu \rho_0(\mu)  \frac{(\mu-k^2)}{(\mu+\mu_0)} \frac{ |T(k)|^2}{\mu+k^2-i \epsilon}  \Bigg]\, .\nn
\ea
Despite being a different quantity, \eqref{TTamp} is conceptually similar to \eqref{KKdispersion} where the functions  $\rho_2(\mu)$ and $\rho_0(\mu)$ are positive by unitary, and are analogous to $\Im[n(\omega)]$, and the subtraction constants $C_2(\mu_0)$ and $C_1(\mu_0)$ analogous to $n(\infty)$. As for the standard K\"allen-Lehmann spectral representation, the momentum space argument is an analytic function of complex momentum squared $s=-k^2+i \epsilon$ up to a pole at $k^2=0$ and a right-hand branch cut\footnote{The branch cut lies on the real axis for $-k^2 \ge 0$ with the physical value understood to be the limit from the upper half complex plane, hence $s=-k^2 + i \epsilon$.}. In defining \eqref{TTamp} we have introduced an arbitrary subtraction scale even though the result is independent of that scale. This is encoded in the renormalization group style equation (see appendix \ref{app:EFT} for derivation)
\be
\mu_0 \frac{\d }{\d \mu_0} C_S(\mu_0) =-\int_0^{\infty} \d \mu \rho_S(\mu) \frac{\mu_0}{(\mu+\mu_0)^2} < 0 \, .
\ee
Although $\mu_0$ is not the sliding scale or cutoff of the usual renormalization group, it encodes the same flow, and we see that unitarity demands positivity of the flow from the UV into the IR, that is
\be
C_S^{\rm IR} > C_S^{\rm UV} \, .
\ee
While this does not constitute a proof, it certainly leads to the expectation that $C_S^{\rm IR} >0$. Attempts to prove this have been given in \cite{Bellazzini:2019xts} for the case of gravity coupled to a Maxwell field via S-matrix positivity arguments\footnote{Since the curvature squared corrections associated with $C_S$ can be removed with field redefinitions, the S-matrix constraints of \cite{Bellazzini:2019xts} are strictly speaking applied to the $F^4$ and $F^2 R$ terms in the Einstein-Maxwell EFT. However, if we take the perspective that the coefficients of these operators are zero before the field redefinition, then positivity implied by the arguments of  \cite{Bellazzini:2019xts} would indeed infer that $C_2^{\rm IR} >0$.
The K\"allen-Lehmann dispersive arguments are clearly weaker than the S-matrix bounds, since the former are sensitive to field frame.}.
 Specifically positivity of $C_2^{\rm IR}$ would follow if the graviton scattering amplitude with the massless $t$-channel pole removed, has a positive second $s$ derivative in the forward scattering limit. Unfortunately   issues with this proof have recently been pointed out in \cite{Alberte:2020jsk}, together with counter-examples which undermine the requirement $C_2^{\rm IR} >0$. \\

What does this have to do with GWs propagating in curved spacetimes? The answer is that the coefficients $C_S^{\rm IR}$ determine precisely the coefficients of the covariant curvature squared terms in the low-energy effective theory, and they in turn determine whether GWs travel super or subluminally. If we begin with the tree level Wilsonian effective action which includes the leading curvature squared terms expected to arise from integrating out loops and higher spin heavy states (see Eq.~\eqref{eq:covariantaction}), the above dispersive arguments enforce positivity of the coefficients of $R^2$ and $R_{\mu\nu}^2 - R^2/3$ or equivalently stated positivity of the Weyl curvature squared term $W_{\mu\nu\rho\sigma}^2$. We may then use the resulting local equations for the low-energy effective theory to determine an effective equation for the propagation of low-energy GWs precisely analogous to \eqref{LEwaveequation2}. By direct analogy we can define the low-energy speed by the refractive index coefficient $n(0)$ that relates the leading `two derivative' part of this effective equation. The central result of \cite{deRham:2019ctd} is that the low-energy refractive index $n(0)$ is determined at leading order by $C_2^{\rm IR}$, and the sign of the latter directly determines the sub- or super- luminality of the low-energy speed of propagation.
In situations where the leading curvature squared terms vanish, such as Schwarzschild spacetime, it is the curvature cubed terms which determine the leading effect, as calculated in \cite{deRham:2019ctd} and considered in detail in \cite{deRham:2020ejn}.

\section{Dealing with Superluminality in Gravitational Setups}
\label{sec:decouplingLim}

\subsection{Non-Gravitational Criterion}
\label{subsec:nonGrav}

Before getting to our main discussion on the speed of GWs we will review here several key points that arise when dealing with EFTs on a background that spontaneously break Lorentz invariance.
Indeed, it is straightforward to write down Lorentz invariant EFTs which exhibit superluminal propagation around spontaneously broken Lorentz backgrounds. The canonical example is that of a $P(X)$ model (ignoring gravity for now and setting the field on Minkowski), for which the non-minimal kinetic term takes the form  \cite{Aharonov:1969vu}
\ba
\label{eq:PX}
{\cal L} = P(X)= X + \frac{a}{M^4} X^2+\mathcal{O}(X^3/M^8)
\ea
with $X= - \frac{1}{2} (\partial \phi)^2$. For the `wrong' sign choice $a<0$, this leads to superluminal propagation around a simple time-dependent  background\footnote{The same EFT with $a<0$ also leads to superluminal propagation around any other background that spontaneously breaks Lorentz invariance by picking up a preferred direction for $\p_\mu \phi$ no matter whether $\p_\mu\phi$ is timelike or spacelike.} $\phi = \phi(t)$. Indeed the sound speed about such a background is given by
\ba
\label{eq:csPX2}
c_s^2=1-4a \frac{\dot \phi^2}{M^4}+\mathcal{O}\(\frac{\dot \phi^4}{M^8}\)\,,
\ea
and is superluminal for $a<0$.
This is connected with a violation of positivity bounds \cite{Pham:1985cr,Ananthanarayan:1994hf,Adams:2006sv}. \\

The departure of the speed from unity is always suppressed by factors of $\dot \phi^2/M^4$ and is therefore always small within the regime of validity of the EFT. Nevertheless,  even such a small correction can lead to significant macroscopic consequences. This is because in this field theory on Minkowski spacetime setting, there is no limit to how long we may wait to integrate this effect up, and even a small local effect can therefore build up to a macroscopic size. Roughly speaking, at a time $t$, the future lightcone emanating from a given spacetime point at $t_0$ will be larger by a radius
\be
\Delta r = \int^t_{t_0} \d t \, |c_s-1| \sim \frac{2(-a) \dot \phi^2 }{M^4} |t-t_0| \, ,
\ee
and the smallness of $\dot \phi/M^2$ may easily be compensated for by the largeness of $|t-t_0|$ which is otherwise unrestricted. Since this distance will be macroscopically observable, which means not only will it satisfy  $\Delta r \gg M^{-1}$, crucially it is resolvable $\Delta r \gg \lambda$ where $\lambda$ is the wavelength of the propagating fluctuation, we are in a situation where we may imagine violations of causality. At the very least, this would imply a violation of causality as implied by the (asymptotic) Lorentz invariant lightcone. \\

Indeed stated this way, it is clear that  {\bf{any EFT in Minkowski spacetime}} that leads to {\it any} amount of homogenous superluminality, no matter how small, will necessarily lead to macroscopic causality violation after a long enough time. Since there is no restriction on how long we can wait, then we must conclude that standard relativistic causality imposes the strict requirement that\footnote{This is more subtle for inhomogeneous configurations where local time advances can be cancelled by neighbouring local time delays to lead to a net time delay.}
\be
\label{eq:csm1}
c_s \le 1 \, .
\ee
Our central claim is that the same argument {\it does not} apply to the low-energy speed in a curved spacetime or gravitational setting for a number of reasons. As we have already alluded to, in  gravitational theory (in a setup that spontaneously breaks Lorentz invariance), the question is more subtle for two reasons: first, the notion of low-energy speed (i.e. any speed inferred from a low-energy effective theory) is not frame independent and second, due to the nature of the spacetime in question, it may not be possible to integrate up a small departure in speed to make a macroscopic effect. In other words, the would--be superluminality may not be resolvable. When this is the case we cannot necessarily legitimately conclude that there is any causality violation in the regime of validity of the effective theory. In fact we will see that in the two most interesting curved spacetimes, namely FLRW and Schwarzschild, this is exactly what happens. \\

When computed from the leading terms in a low-energy EFT, the low-energy speed may not manifest any explicit frequency dependence. This occurs in particular if only at most second order derivative terms have been included in the low-energy EFT and any other higher derivative term has been considered as irrelevant, as was the case in the $P(X)$ EFT considered previously in \eqref{eq:PX}. We stress however that this is always an artifact of the truncation to the leading EFT interactions. However from the very definition of a low-energy EFT, \eqref{eq:csPX2} can only give an appropriate description for $c_s$ at low-energies, in this case at frequencies of at most $\omega \ll M^2/|\dot \phi|$. Beyond the regime of validity of the low-energy EFT, the computation of the speed is simply not valid. Even  if the formula does not appear to break down mathematically at that order, it does break down physically when flowing to higher energy as higher order irrelevant operators ought to be taken into account, until ultimately the theory ought to be traded for its higher energy counterpart, explicitly including heavier modes. See Ref.~\cite{deRham:2018red} for a discussion.

\subsection{Decoupling Limit}
\label{sec:DL}

In many situations, in a given gravitational effective field theory, it may be possible to take a decoupling limit\footnote{
A decoupling limit field theory should not be confused with the original Lagrangian simply evaluated on a Minkowski background.
See for instance Refs.~\cite{deRham:2014zqa,deRham:2018dqm} for discussions on the physical meaning of a decoupling limit.} $\mpl \rightarrow \infty$, keeping some other interaction scale $M$ in the system fixed, for which the helicity-2 gravitational degrees of freedom decouple from all other degrees of freedom while maintaining the interactions that arise at the lowest energy scale $M$. Whenever this is possible, the resulting decoupled effective theory can be analyzed from the perspective of an interacting field theory on a fixed (Minkowski or other curved) background, and in this situation the above argument \eqref{eq:csm1} is expected to be valid. With this in mind we may then declare that if {\it in the field frame for which the decoupling limit is well-defined}
\be\label{eq:speed1}
\lim_{\substack{\mpl \rightarrow \infty, \\ \text{fixed } M}}
c_s \le 1 \, ,
\ee
for all species then causality is (expected to be) satisfied.
The condition on the {\it field frame} is crucial as we will see in Section~\ref{fieldframeexample}. More precisely we will see that rate to which the effect goes to zero as $\mpl \rightarrow \infty$ is crucial, and a more refined version of this statement is the bound \eqref{bound4}.

\paragraph{$\bullet$  Superluminal Speed in the Decoupling Limit.--}
As a first example, we may consider the gravitational version of our canonical example in the previous subsection~\ref{subsec:nonGrav}. Promoting the previous example to a gravitational effective field theory
\be\label{EFT1}
{\cal L} = \sqrt{-g} \left ( \frac{\mpl^2}{2}R +X + \frac{a}{M^4} X^2   \right)\,,
\ee
it is straightforward to take the limit $\mpl \rightarrow \infty$ with $g_{\mu \nu} = \eta_{\mu\nu}+ h_{\mu\nu}/\mpl$ keeping the scale $M$ fixed. In this limit, we are left with two decoupled sectors, on one side a free massless spin-2 degree  of freedom $h_{\mu\nu}$ and on the other side an  interacting $P(X)$ scalar field theory on Minkowski identical to that considered in \ref{subsec:nonGrav} to which the usual superluminality and positivity bounds violation arguments apply.

\paragraph{$\bullet$  Luminal Speed in the Decoupling Limit.--}
On the other hand, we may now consider a modification to the low-energy speed that is parametrically suppressed by powers of $M^2/\mpl^2$ relative to the previous effect, take for instance
\be\label{eq:speed2}
c_s^2 = 1 + \frac{|a| \dot \phi^2}{\mpl^2 M^2} + \dots\,.
\ee
As we shall see, this is closer to the typical situation for the low-energy speed of GWs with corrections as in \cite{deRham:2019ctd}. In this case it is not possible to take the limit $\mpl \rightarrow \infty$ and have $c_s$ differ from unity without something else blowing up. For instance we may try to scale $M \sim 1/\mpl$ but with the understanding that $M$ sets the scale of other irrelevant operators in the effective theory, this would inevitably lead to a break down of the low-energy EFT at arbitrarily low scales, invalidating the calculation of the speed.\\

Our central claim is that whenever a situation like \eqref{eq:speed2} occurs where $M$ is related to the cutoff of the low-energy EFT in which the speed has been computed (or to the scale of irrelevant operators), then the condition \eqref{eq:speed1} is actually satisfied and it would then not be legitimate to demand that $c_s \le 1$ for all degrees of freedom away from the decoupling limit (i.e. at finite $\mpl$). Specifically we will see that in actual examples, the low-energy speed $c_s$ is typically expected to be superluminal without leading to any macroscopic violation of causality. The key to this is the smallness of the effect, and this is in turn tied to the fact that the speed is luminal in the limit $\mpl \rightarrow \infty$. As long as \eqref{eq:speed1} continues to hold, we do not anticipate any violation of causality.
More precisely we shall see that if
\be \label{bound4}
\lim_{\mpl \rightarrow \infty} |c_s^2-1| \sim \mpl^{-\alpha}\,,
\ee
again {\it in the field frame for which the decoupling limit is well-defined},
with $\alpha \ge 2$, there will be no macroscopic observable effects. A single $\mpl$ suppression would not be sufficient, however all gravitational induced corrections to the sound speed
arise at a minimum with a $\mpl^2$ suppression (in a local theory).

\subsection{Macroscopic Superluminality allowed by Analyticity}\label{fieldframeexample}

In order to illustrate the subtleties that emerge when dealing with superluminalities in a gravitational theory, we give here an example of an effective theory for which superluminal GWs are required by analyticity! Consider the effective Lagrangian of the form
\be\label{EFT2}
{\cal L}_{\rm EFT} = \sqrt{-g} \left(  \frac{\mpl^2}{2} R + a \frac{\mpl^2}{2M^4} G^{\mu\nu} \nabla_{\mu} \phi \nabla_{\nu} \phi - \frac{1}{2} (\nabla \phi)^2  + \dots\, \),
\ee
which includes a non-minimal coupling between gravity and the scalar. It is straightforward to show that on considering perturbations around a cosmological solution sourced by a time dependent scalar $\phi(t)$, the GWs propagate superluminally if $a>0$ and subluminally if $a<0$. Furthermore this effect is a macroscopically observable one since
\be \label{speedcorrection}
\Delta c_s^2 \sim a \frac{\dot \phi^2}{M^4} \, .
\ee
Surely then, since this superluminality is macroscopically observable, causality/analyticity considerations will demand that $a<0$? In fact it is straightforward to see that this is not the case and the precise opposite actually holds. Indeed, one can change frame so that the Lagrangian \eqref{EFT2} exactly matches  \eqref{EFT1} by the simple redefinition
\be\label{fieldred}
g_{\mu\nu} \rightarrow g_{\mu\nu} + a \frac{1}{2M^4}  \nabla_{\mu} \phi \nabla_{\nu} \phi \, ,
\ee
whence standard positivity bounds following from analyticity impose that $a>0$ as the `causal' choice.
In terms of the canonical normalized gravitational fluctuations $g_{\mu\nu} = \eta_{\mu\nu} + h_{\mu\nu}/\mpl$ this is
\be\label{fieldred2}
h_{\mu\nu} \rightarrow \tilde h\mn = h_{\mu\nu} + a \frac{\mpl}{2M^4}  \nabla_{\mu} \phi \nabla_{\nu} \phi \, .
\ee
This example nicely illustrates two points (a) the ambiguity of speed under field redefinitions and (b) the importance of causality constraints being imposed in the frame in which the decoupling limit is well defined. Indeed, unlike the field frame implicit in \eqref{EFT1} for which the decoupling limit $\mpl \rightarrow \infty$ is well defined, the Lagrangian \eqref{EFT2} does not have a well defined decoupling limit. Indeed in taking the limit $\mpl \rightarrow \infty$ in this frame we would have
\be
\label{eq:FrameDL}
\lim_{\mpl \rightarrow \infty }\frac{\mpl^2}{M^4} G^{\mu\nu} \nabla_{\mu} \phi \nabla_{\nu} \phi \sim \frac{\mpl}{M^4} \partial \partial h  \partial \phi \partial \phi \rightarrow \infty \, .
\ee
Stated equivalently, the field redefinition \eqref{fieldred2} blows up in the decoupling limit explaining the inequivalence of the two frames. We can make \eqref{fieldred2} and \eqref{eq:FrameDL} finite in the limit $\mpl \rightarrow \infty$ by scaling $M =\mpl^{1/4} (\tilde \Lambda)^{3/4}$ keeping $\tilde \Lambda$ fixed, but then the effect \eqref{speedcorrection} vanishes as $\mpl \rightarrow \infty$. \\

The lesson to learn from this is that simply demanding that for a given EFT the speed of propagation of fields is (sub)luminal in a given field frame is not only unjustified, it may even explicitly violate the requirements that do come from causality. For this reason, the only safe requirement to impose on the EFT is that given in \eqref{eq:speed1} which only applies in a field frame for which a decoupling limit exists.

\section{Cosmology in the EFT of Gravity}
\label{sec:cosmology}

\subsection{Speed of Gravity in Cosmology}

Let us now focus our discussion on the specific case of cosmological spacetimes. As reviewed in appendix~\ref{app:EFT}, the leading corrections to the low-energy EFT for gravity may be expressed in the form
\be\label{eq:covariantaction2}
{\cal L}^{\rm EFT} = \sqrt{-g} \left( \frac{\mpl^2}{2} R + C_{R^2}^{\rm IR} R^2   +C_{W^2}^{\rm IR} W_{\mu\nu\alpha\rho}^2 + C_{\rm GB} {\rm GB}  \right) + \text{higher derivative terms} \, ,
\ee
where ${\rm GB}$ designates the Gauss-Bonnet term and $W$ the Weyl tensor.
In addition when matter is included, we may allow for non-minimal matter curvature interactions, as for example $R F^2$ terms in the case of Einstein-Maxwell. In order to focus on the genuine gravitational interactions, we shall not consider these non-minimal matter interactions in what follows (including them can lead to additional sources of superluminalities that can be dealt with in a more standard way). Within the low-energy EFT, one cannot determine the sign of the coefficients $C_{R^2, W^2}^{\rm IR}$ but
as argued in appendix~\ref{app:EFT} the positivity of the RG flow \eqref{flow} implies
\be
C_{W^2}^{\rm IR} > C_{W^2}^{\rm UV} \,,  \qquad  C_{R^2}^{\rm IR} > C_{R^2}^{\rm UV}  \, ,
\ee
and in what follows we shall make the a priori not-so-unreasonable assumption that $C_{W^2}^{\rm IR}$ may be positive (see Ref.~\cite{Alberte:2020jsk} for a more precise discussion and potential caveats).\\

Given a covariant action of the local form \eqref{eq:covariantaction2} encoding the EFT corrections, it is straightforward to compute the corrections to the equation of motion for tensor fluctuations on a cosmological background as done in \cite{deRham:2019ctd}. Identifying what we mean by speed is however slightly subtle since the truncated equation of motion contains higher time derivatives. These may be removed with field redefinitions and traded for space derivatives just as in the conversion from \eqref{LEwaveequation} to \eqref{LEwaveequation2}. After this is done, the equation of motion for tensor GWs on FLRW may, by virtue of symmetry, be put in the following form
\be \label{eq:exacteqn}
\partial_{\eta}^2 h = -\sum_{n=0}^{\infty} \beta_n(\eta) k^{2n} h \, ,
\ee
where we work in  conformal time $\d s^2 = a(\eta)^2(- \d \eta^2 + \d \vec x{\, }^2)$. The previous relation  may be rearranged and express in the form
\be\label{eq:exacteqn2}
\partial_{\eta}^2 h +m_{\rm eff}^2 h + \tilde c_s^2(k,\eta) k^2 h =0 \, ,
\ee
where $m_{\rm eff}^2=\beta_0(\eta)$ is an effective mass and $\tilde c_s^2(k,\eta) = \sum_{n=1}^{\infty} \beta_n(\eta) k^{2n-2}$ an effective $k$ dependent sound speed. We then define the low-energy sound speed to be $c_s^2(\eta)\equiv \tilde c_s^2(0,\eta) =\beta_1(\eta)$, namely the speed of propagation implied by the truncated equation
\be\label{eq:lowenergyeqn}
\partial_{\eta}^2 h +m_{\rm eff}^2 h + c_s^2(\eta) k^2 h =0 \, .
\ee
Explicit calculation using \eqref{eq:covariantaction} gives the {\it low-energy} speed \cite{deRham:2019ctd}
\be\label{eq:finalspeed}
c_s^2 = 1- \frac{16 C_{W^2} \dot H}{\mpl^2} \, .
\ee
Since the null energy condition requires $\dot H<0$, then $C_{W^2}>0$ would imply that this low-energy speed is slightly superluminal. Note that we do not include the effective background-generated mass $m_{\rm eff}$  in consideration of the speed of propagation, because what is relevant is the causal support of the retarded propagator. If \eqref{eq:lowenergyeqn} were the exact equation this would be determined by $c_s^2(\eta)$ alone \cite{Caldwell:1993xw}. This is in agreement with what is typically meant by the low-energy speed. If the retarded propagator for the exact equation \eqref{eq:exacteqn} is determined as a perturbative expansion in spatial derivatives, with the (up to) two derivative terms \eqref{eq:lowenergyeqn} taken as the leading part, then at any finite order in perturbations the causal support for the retarded propagator will be determined by \eqref{eq:lowenergyeqn}. Clearly the relevant question is, when is this a good indication of the true causal support of the exact retarded propagator.

\subsection{Validity of EFT in FLRW}\label{validity}

Before proceeding, we need to address the conditions for the validity of the EFT, e.g. the validity of equation \eqref{eq:exacteqn} to describe the evolution of GWs. In the application of the Wilsonian effective theory with cutoff $M$ we can only describe momenta for which covariant operators are small relative to the cutoff scale, e.g. $\Box \ll M^2$ and $R \ll M^2$ . In the cosmological context, since Lorentz invariance is broken by the background, this means we can only use the effective theory to describe the evolution of modes in the region where
\be\label{eq:EFTbound}
| \dot H(t) |\frac{| \vec k |^2}{a(t)^2} \ll M^4 \, ,
\ee
i.e. $k/a(t) \ll \Lambda_c(t)$ where $\Lambda_c = M^2/\sqrt{-\dot H}$. In the typical situation in which $\dot H \sim {\cal O}(H^2)$ this may be stated as
\be\label{eq:EFTbound2}
\frac{| H(t) ||\vec k |}{a(t)} \ll M^2 \, .
\ee
Note that this scale is much higher\footnote{Assuming the much tighter requirement $k\sim a(t) M$ typically considered in trans-Planckian arguments would only help with our following argumentation.} than that typically considered in trans-Planckian type arguments where it would be argued that the EFT breaks down when $k\sim a(t) M$ \cite{Martin:2000xs,Jacobson:1999zk}. The reason being is that we assume the underlying theory is Lorentz invariant, and so we require a locally Lorentz invariant combination to be comparable to $M^2$. In the cosmological context where only time translations are broken, we may for example decompose the Ricci tensor in the manner
\be
R_{\mu \nu}  = \Omega^2 g_{\mu\nu} + \kappa_{\mu}\kappa_{\nu}\,,
\ee
where $\kappa_{\mu}$ is a non-normalized time-like vector (since we are dealing with cosmology here). Given an on-shell wave of momentum $k_{\mu}$ for which $k_{\mu}k^{\mu} \approx 0$, then \eqref{eq:EFTbound2} is the locally Lorentz invariant bound
\be
|\kappa^{\mu} k_{\mu}| \ll M^2 \, .
\ee
This may be taken together with the requirement that $\kappa_{\mu} \kappa^{\mu} \ll M^2$ and $|\Omega| \ll M$ which require $H^2 \ll M^2$  and $|\dot H| \ll M^2$.
Indeed the argument for why \eqref{eq:EFTbound} is the more general condition and not \eqref{eq:EFTbound2} is that de Sitter invariance in the limit $\dot H \rightarrow 0$ is sufficient to ensure validity of EFT at arbitrarily on-shell high momenta. \\

To clarify this, let us think of a typical example EFT organized in the standard manner where all irrelevant operators are suppressed by the common scale $M$ to the appropriate power. Schematically the effective action takes the form\footnote{This is for example the schematic form of curvature dependence in the low-energy EFT for string theory in which $M$ is the string scale $1/\sqrt{\alpha'}$ which is parametrically below the Planck scale \cite{Gross:1986iv,Metsaev:1986yb}.}
\be\label{EFT100}
S_{\rm EFT} = \frac{\mpl^2}{2}\int \d^4 x  \sqrt{-g} \left( R + M^2 \sum \alpha_{ab}  \( \frac{\rm \nabla}{M} \)^{2a} \( \frac{\rm Riemann}{M^2} \)^b \dots  \right) \, ,
\ee
with the usual understanding that we allow for all local scalar operators constructed out of the appropriate number of powers of the Riemann tensor and covariant derivatives in any order.
Given the underlying locality and Lorentz invariance, any term in the effective dispersion relation for GWs around a curved background not of the local Lorentz invariant form $(\omega^2 - k^2/a^2)$ will necessarily come suppressed by some power of the background curvature quantities $H^2$, $\dot H$ and derivatives thereof. Since these terms spontaneously break Lorentz invariance they may come together with $k^2/a^2$ and $\omega^2$ terms and so will naturally package in dimensionless combinations of the form
\be
 \frac{\dot H k^2/a^2}{M^4} \sim \frac{\dot H \omega^2}{M^4} \, .
\ee
For example these will arise from terms in \eqref{EFT100} with factors of $M^{-4}R^{\mu\nu} \nabla_{\mu} \nabla_{\nu}$. Terms with the same number of powers of curvature, but higher powers of $k$ such as $\frac{\dot H k^4/a^4}{M^6} $ will necessarily only arise is the quasi-Lorentz invariant combination
\be
\frac{\dot H k^2/a^2}{M^6} (\omega^2 - k^2/a^2)  \, ,
\ee
as for example coming from terms like $M^{-6}R^{\mu\nu} \nabla_{\mu} \nabla_{\nu} \Box$ (acting for instance on the scalar curvature). This is in essence due to index contraction, if we limit ourselves to a fixed number of powers of curvature, since Weyl is zero for FLRW, once the Ricci tensor indices have been contracted, all remaining indices must be contracted with the metric which locally takes a Lorentz invariant form. Thus schematically the effective form of the dispersion relation will be (in terms of the physical momentum $\tilde k=k/a$)
\ba
\label{eq:disp}
\omega^2 - \tilde k^2  - m^2_{\rm eff}/a^2
+M^2 \sum \beta_{a_1 \dots a_7} &\Bigg[& \(\frac{ \omega^2 - \tilde k^2}{M^2}  \)^{a_1}   \(\frac{1}{M^2} \frac{\d^2}{\d t^2} \)^{a_2}   \(\frac{ H}{M^2} \frac{\d}{\d t} \)^{a_3}\\
&&  \(\frac{ H^2}{M^2}  \)^{a_4} \(\frac{ \dot H}{M^2}  \)^{a_5}  \( \frac{\dot H \tilde k^2}{M^4}\)^{a_6}\( \frac{\dot H  \omega^2}{M^4}\)^{a_7} \Bigg]=0\, .\nn
\ea
The condition that the EFT remains under control requires that at a minimum the $\beta$ corrections are small or more precisely that the $\beta$ series is at least convergent in an asymptotic series sense. Since we are allowed arbitrary integer powers of the $a_i$, this will only be true if each of the dimensionless ratios in brackets are kept smaller than unity.
Hence in addition to the expected requirements that the curvature remains small, $H^2 \ll M^2$  and $|\dot H| \ll M^2$,  we infer that
\be
\frac{|\dot H |\tilde k^2}{M^4}= \frac{|\dot H | k^2/a^2}{M^4} \ll 1 \, .
\ee
This implies that in the typical situation for which $\dot H \sim {\cal O}(H^2)$, the momentum cutoff appropriate for an on-shell state, i.e. a propagating gravitational wave, is therefore as specified in \eqref{eq:EFTbound2}. Due to redshifting in the cosmological context of an expanding Universe, the bound \eqref{eq:EFTbound} is strongest at the earliest times which is where we shall make use of it.

\subsection{Causality constraint}

While it is known and observed that in many media the low-energy phase and group velocities may temporarily become superluminal, this is only in conflict with causality if the superluminality may be integrated up to a macroscopic effect for which the lightcone is clearly larger than the Lorentzian lightcone. One way to characterise this in asymptotically flat spacetimes is to ask whether there is an `asymptotic superluminality' \cite{Gao:2000ga,Camanho:2014apa}. In practice, this amounts to asking whether there can be an integrated time advance in a scattering event, which would imply that the signal from a scattering process could arrive before that of an unscattered wave  -- in a Lorentz invariant theory this would then be associated to some type of acausality. In the cosmological context (or any other curved geometry which is not asymptotically flat), we do not have such a clean tool, and any S-matrix calculation of this form would only be approximately valid at subhorizon scales. We can however ask, by virtue of the symmetry of the FLRW spacetime, how much larger the lightcone of propagation is emanating from some event after many Hubble times. On first sight, we may imagine that even the tiniest amount of superluminality in the low-energy phase could be integrated up to some large observable effect over the entire age (or even future) of the Universe. Crucially, this is not the case as we now explain. \\

Let us work with the effective metric seen by GWs in the EFT of gravity, i.e. an effective metric with speed $c_s(t)$ as in \eqref{eq:finalspeed}. We now consider the future lightcone emanating from a spacetime event at time $t_i$ as determined with respect to this effective metric. If $C_{W^2}>0$ then at a given time $t>t_i$ this lightcone is larger than the usual FLRW lightcone by a radial distance $\Delta r$
\be
\Delta r(t)  = a(t)  \int_{t_i}^t \frac{\d t'}{a(t')} (c_s(t')-1)\,.
\ee
Using \eqref{eq:finalspeed} at leading order in the EFT expansion this distance is
\be
\Delta r(t)  = a(t)  \int_{t_i}^t \frac{\d t'}{a(t')} \left( - \frac{8 C_{W^2} \dot H(t')}{\mpl^2}  \right) + \dots\,.
\ee
In an expanding Universe, the integrand on the right hand side is bounded by
\be
\Delta r(t)  \le \frac{a(t)}{a(t_i)}  \int_{t_i}^t \d t' \left( - \frac{8 C_{W^2} \dot H(t')}{\mpl^2}  \right) + \dots\,.
\ee
This implies
\be
\Delta r(t)  \le \frac{8 C_{W^2} }{\mpl^2}  \frac{a(t)}{a(t_i)} (H(t_i) - H(t)) \le \frac{8 C_{W^2} }{\mpl^2}  \frac{a(t)}{a(t_i)} H(t_i) \,,
\ee
given $H(t)< H(t_i)$ for $t>t_i$, assuming the null energy condition is satisfied.

\paragraph{Post--inflation period.--}
For a mode of a given comoving momentum $k$, the earliest time at which we can trust the EFT calculation of the speed is set by \eqref{eq:EFTbound2} to be such that (assuming that for most of cosmic history $\dot H \sim {\cal O}(H^2)$ which is true post inflation)
\be
\frac{H(t_i)}{a(t_i)} \ll \frac{M^2}{|\vec k|} \, ,
\ee
from which we infer
\be
\Delta r(t)  \le \frac{8 C_{W^2} }{\mpl^2}   \frac{a(t)}{k} \ll \frac{4 C_{W^2}  M^2}{ \pi \mpl^2} \lambda(t)\,,
\ee
where $\lambda(t) = 2 \pi a(t)/k$ is the physical wavelength. Finally the cutoff of the EFT should be at most $M^2 \lesssim \mpl^2/C_{W^2}$, so the bound essentially becomes
\be
\label{eq:Delta_r}
\Delta r(t)  \ll \lambda(t) \, .
\ee
We recall that $\Delta r$ represents the distance that low-energy GWs may propagate outside the light cone set by the FLRW background metric (i.e. the light cone seen by minimally coupled fields) if $C_{W^2}>0$. For any GW  this distance is always much less than the actual physical wavelength $\lambda$ of the GW (if we ensure that we remain within the regime of validity of the low-energy EFT). This distance is therefore not resolvable, and if it ever were resolvable, one would not be able to trust the result as it would rely on applying the low-energy EFT beyond its regime of validity. Thus causality remains intact provided we limit ourselves to asking questions that are fully within the regime of validity of the EFT.

\paragraph{Quasi--inflationary period.--}
The situation is only slightly more subtle when there is a quasi-inflationary (or late--time acceleration) period for which $|\dot H| \ll H^2$. Consider for example a constant equation of state $w \approx -1$ for which the scale factor takes the form $a(t)=a(t_b) (t/t_b)^{2/(3 (1+\omega)}$. The additional contribution to the comoving displacement coming from an inflationary epoch $t_e>t>t_b$ is
\ba
 \int_{t_b}^{t_e} \frac{\d t'}{a(t')} \left( - \frac{8 C_{W^2} \dot H}{\mpl^2}  \right) &=& \frac{8 C_{W^2}}{\mpl^2} \frac{1}{(1+\frac{3}{2}(1+\omega))}\frac{1}{a(t_b) t_b}  \( 1- \(\frac{t_b}{t_e}\)^{\frac{5+3\omega}{(3 (1+\omega)}}\) \nn \\
 & \lesssim& \frac{8 C_{W^2}}{\mpl^2} \frac{\sqrt{3(1+\omega)/2}}{(1+\frac{3}{2}(1+\omega))}\frac{\sqrt{-\dot H(t_b)}}{a(t_b) } \, .
\ea
This is further suppressed by $\sqrt{1+\omega}$ relative to the previous estimate, and so on applying the condition \eqref{eq:EFTbound} at the beginning of inflation we are led to the same conclusion \eqref{eq:Delta_r}.

\subsection{Suppression is key}\label{suppression}
The key as to why causality is not being violated by the superluminal speed here is the smallness ($\dot H/\mpl^2$) of the effect, i.e. the gravitational suppression. To clarify this let us return to the case of a genuinely acausal example as in the  `wrong' sign $P(X)$ model with $a=-|a|$ \cite{Aharonov:1969vu}, i.e.
\be
{\cal L} = - \frac{1}{2} ( \partial \phi)^2 - \frac{|a|}{4M^4} (\partial \phi)^4 -V(\phi)\, ,
\ee
then the speed of propagation for the scalar about say a time-dependent background $\phi(t)$ is given in \eqref{eq:csPX2}. \\

Although the departure from unity for the speed is small in an EFT sense, its macroscopic secular effect can be arbitrary large, even on an FLRW background. To illustrate this, let us suppose that this scalar is also the dominant source for the background expansion, then the leading order Raychaudhuri equation is $\dot H = -\dot \phi^2/(2 \mpl^2) + \dots$ and so
\be
\label{eq:csPX}
c_s^2 = 1 + 8 |a| \frac{(-\dot H) \mpl^2}{M^4} \, .
\ee
Assuming order unity Wilson coefficient $a \sim \mathcal{O}(1)$, the departure from luminality is larger by a factor of $\mpl^2/M^2$ compared to the previous example. We infer a maximal displacement of the light cone of order
\be\label{Badcase}
\Delta r(t)  \sim \frac{\mpl^2}{M^2} \lambda(t) \, .
\ee
Whenever $M \ll \mpl$ we can engineer a situation where there is an observable violation for causality with $\Delta r(t) \gg \lambda(t)$, justifying the inherent acausality of the wrong sign $P(X)$ model.  By comparison, in the low-energy EFT for gravity, the corrections to the speed is suppressed by an additional factor of $M^2/\mpl^2 $ (or in the BH case of section \ref{sec:BH} by a factor $(M/\mpl)^4$) relative to \eqref{eq:csPX}, which is precisely what makes the displacement unobservable. It is clear from \eqref{Badcase} that we need at least two powers of $\mpl$ suppression to ensure the unobservability of this effect \eqref{bound4}, justifying the claim made in \eqref{bound4}.

\subsection{Secular effects}

To restate the previous results slightly differently, suppose we tried to infer the retarded Green's function describing the response of a field to a source. From the exact equation \eqref{eq:exacteqn2} the momentum space retarded Green's function may be defined by
\be\label{Greensdef}
\left[ \partial_{\eta}^2  +m_{\rm eff}^2  + c_s^2(k,\eta) k^2  \right] G_{\rm ret}(\eta,\eta') =\delta(\eta-\eta')  \, .
\ee
Ideally this equation would be solved exactly, however we only know its form within the context of an EFT expansion. The picture closest to the classical one is where we infer the propagator by means of a WKB approximation as discussed for example in \cite{Bunch:1979uk}. The exact retarded propagator is given by
\be
G_{\rm ret}(\eta,\eta') = \theta(\eta-\eta')  i   \left[  h_k(\eta) h_k^*(\eta') -  h_k(\eta') h_k^*(\eta) \right] \, ,
\ee
where $h_k(\eta)$ are the normalized `positive frequency' solutions of \eqref{eq:exacteqn2}. If we implicitly resum the secular contribution from the sound speed, these will be build out of modes of the WKB form
\ba\label{WKB1}
h_k(\eta)  &\sim& \frac{1}{\sqrt{2 \omega_k(\eta)}} e^{ \mp i  k \int^{\eta} \d \eta' c_s(\eta') } \\
&\sim&  \frac{1}{\sqrt{2 \omega_k}} e^{ \mp i k  \int^t \d t' \frac{c_s(t')}{a(t')} } =  \frac{e^{\mp i k  \int^t  \frac{\d t'}{a(t')}}}{\sqrt{2 \omega_k}} \left[e^{\mp i k  \int^{t} \d t' \frac{c_s(t')-1}{a(t')} } \right] \, .\nn
\ea
The resulting propagator will have causal support on the lightcone determined by the speed $c_s$ in the exponent of the exponentials. However, the secular resummation implicit in \eqref{WKB1} only makes sense if the argument of the exponential in square brackets becomes of order unity or larger, otherwise this effect is clearly a perturbative one.
However, as we have seen previously, provided we demand the EFT bound \eqref{eq:EFTbound} then
\be\label{secular}
\Bigg|  i k  \int^{t} \d t' \frac{c_s(t')-1}{a(t')} \Bigg| \ll 1\,,
\ee
and so in the calculation of the Green's function we may always treat the exponential perturbatively
\be \label{WKB10}
h_k(\eta)  \sim   \frac{e^{\mp i k  \int^t  \frac{\d t'}{a(t')}}}{\sqrt{2 \omega_k}} \left[1 \mp i k  \int^t \d t' \frac{c_s(t')-1}{a(t')} + \dots \right] \,.
\ee
When computed in this manner, the resulting Green's function will have at any finite order the same lightcone structure as the FLRW background metric as determined by the leading exponential $e^{i \vec k . \vec x \mp i k  \int^t \frac{\d t'}{a(t')} }$. We can only justify the resummation of the terms that arise from expanding the exponential if those were  the only terms to arise from the EFT expansion. But this is of course not the case, they represent only a subset of contributions and since their individual contribution remains perturbative we have no reason to expect that for example the term of quadratic order in $( i k  \int^t \d t' \frac{c_s(t')-1}{a(t')} )$ is any larger than other term that arise at the same order in the EFT expansion. \\

The implication of \eqref{eq:EFTbound}  is that since \eqref{secular} is satisfied, it implies that \eqref{Greensdef} is best solved as a perturbation series defined by iterating the equation
\be
\label{eq:GretPert}
\left[ \partial_{\eta}^2  +m_{\rm eff}^2  +k^2  \right] G_{\rm ret}(\eta,\eta') =\delta(\eta-\eta') -(c_s^2(k,\eta) -1) k^2  G_{\rm ret}(\eta,\eta')  \,
\ee
that is
\be
G_{\rm ret}(\eta,\eta') = G^0_{\rm ret}(\eta,\eta') - \int^{\eta}_{\eta'} \d \eta''  G^0_{\rm ret}(\eta,\eta'')(c_s^2(k,\eta'') -1) k^2   G^0_{\rm ret}(\eta'',\eta')  + \dots \, ,
\ee
for which $\left[ \partial_{\eta}^2  +m_{\rm eff}^2  +k^2  \right] G^0_{\rm ret}(\eta,\eta') =\delta(\eta-\eta') $ has the causal support of the background metric.
This is legitimate as long as the there is {\bf no secular growth} in the perturbative expansion, which amounts to the requirement \eqref{secular} which in the EFT of gravity follows from \eqref{eq:EFTbound}.
We see that the EFT validity condition \eqref{eq:EFTbound} is crucial to understanding how causality is preserved. It is the presence or absence of secular growth in the perturbative expansion that tells us whether or not the sound speed departure from unity is physical or not.

\section{Black Holes in the EFT of Gravity}
\label{sec:BH}

\subsection{Speed of Gravity near Black Holes in the EFT of Gravity}

As a second class of configurations that spontaneously breaks Lorentz invariance, we consider $D=4$ BH types of solutions and focus on static and spherically symmetric Ricci-flat vacuum configurations. This situation is not only particulary interesting phenomenologically, it also provides an explicit asymptotically flat example where S-matrix arguments can be applied. In this vacuum flat case the $R^2$ operators in the EFT of gravity affect neither the background solution nor the propagation of GWs to first order in curvature corrections. Instead, the leading contributions arising from the dimension-six operators of the form
\ba
\label{eq:l3}
\mathcal{L}_{\rm D6}=\frac{1}{M^2}&\Bigg[& d_1 R\Box R + d_2 R_{\mu \nu} \Box R^{\mu \nu} +d_3 R^3 + d_4 R R_{\mu \nu}  \\
&+& d_5 R R_{\mu \nu \alpha\beta} + d_6 R_{\mu \nu}^3 + d_7 R^{\mu \nu} R^{\alpha \beta}R_{\mu\nu\alpha\beta} + d_8 R^{\mu\nu} R_{\mu\alpha \beta \gamma} {R_{\nu}}^{\alpha \beta \gamma} \nonumber \\
&+&  d_9 \tensor{R}{_\mu_\nu^\alpha^\beta} \tensor{R}{_\alpha_\beta^\gamma^\sigma} \tensor{R}{_\gamma_\sigma^\mu^\nu}+ d_{10}\tensor{R}{_\mu^\alpha_\nu^\beta}\tensor{R}{_\alpha^\gamma_\beta^\sigma} \tensor{R}{_\gamma^\mu_\sigma^\nu}\Bigg],\nonumber
\ea
where $M$ is the `naive' cutoff of the EFT. We consider the background solution to be that solution which is Schwarzschild if the $R^3$ operators were absent, i.e. that of a corrected non-rotating black hole. In Schwarzschild coordinates,
the equation of motion for the odd and even polarizations of the GWs $h$ is the same and governed by an effective metric $Z\mn$ \cite{deRham:2020ejn}
\ba\label{eq:scalar}
Z^{\mu\nu} \D_\mu \D_\nu h + V h = 0,
\ea
where the effective metric is expressed as
\ba\label{Effmetric1}
Z\mn \d x^\mu \d x^\mu=- Z_t\, \d t^2+Z_r^{-1}\d r^2+  Z_\Omega\, r^2\, \d \Omega^2\,,
\ea
and the corrected metric functions are
\ba
Z_t &=& Z_r =
 1-\frac{r_g}{r}
 + \frac{r_g}{2 M^2\mpl^2 r^5}\Bigg[6 (4 d_5+d_8)  \left(5 \frac{r_g}{r}-4 \frac{r_g^2}{r^2}\right)\\
 &-&4 d_9 \left(72 -171 \frac{ r_g}{r}+94 \frac{r_g^2}{r^2}\right) -d_{10} \left(144 -297 \frac{r_g}{r}+152 \frac{r_g^2}{r^2}\right)\Bigg]+\mathcal{O}\(M \mpl r_g^2\)^{-4}\nonumber\,,
\ea
and
\ba
Z_\Omega=1+72 \frac{r_g}{M^2 \mpl^2 r^5} (2 d_9+d_{10})\(1-\frac{r_g}{r}\)+\mathcal{O}\(M \mpl r_g^2\)^{-4}\,.
\ea
Here $r_g$ is the Schwarzschild radius of the BH solution in GR without the corrections from the EFT. In this EFT, the BH horizon $r_g$ is slightly displaced by an amount proportional to $(M \mpl r_g^2)^{-2}\ll 1$ and is necessarily the same for every species, no matter how they couple to gravity. This is likely linked to the horizon theorem proven in \cite{Shore:1995fz} within the context of QED (see also \cite{Shore:2000bs,Hollowood:2009qz}). \\

We can show that on the background of BH--like solutions, the speed of GWs can be both superluminal or subluminal depending on the signs of the EFT of gravity. Typically when the angular speed is subluminal, the radial speed is superluminal, as is the case if we think of this EFT as arising from integrating out a spin-1/2 field. Generically, the radial speed is given in terms of the coupling constants $d_{9,10}$ as follows \cite{deRham:2020ejn}
\begin{eqnarray}
\label{eq:cs}
&& c_s^2(r) = 1 +  \, \Delta c_s^2(r)\,, \\
&& \quad {\rm with} \quad \Delta c_s^2(r) =-144 \frac{2 d_9 + d_{10}}{M^2\mpl^2 r_g^4} \(1-\frac{r_g}{r}\)\frac{r_g^5}{r^5}+\mathcal{O}\( \(M \mpl r_g^2\)^{-4}\)\,, \nn
\end{eqnarray}
and is superluminal whenever $2d_9+d_{10}<0$. In the past, these types of arguments have been used to constraint EFTs. We emphasize here that this would be the wrong approach. First the choice $2d_{9}+d_{10}>0$ would still lead to superluminalities in other configurations (e.g. superluminal angular low-energy speed), and second as in the case of the EFT of gravity in cosmological settings, the amount of low-energy superluminality is so suppressed that it can never lead to any macroscopic violation of causality as we explain below.

\subsection{Validity of EFT in BH spacetime}\label{validity2}

The discussion of the validity of the EFT in a BH spacetime is closely analogous to that in FLRW in section~\ref{validity}. One crucial difference however is that the leading order background geometry has vanishing Ricci tensor. Thus corrections to the propagation equations will be governed by the Weyl tensor and its covariant derivatives. Given an on--shell mode of momentum $k_{\mu}$ with $k_{\mu}k^{\mu} \approx 0$, the naive highest order in $k_{\mu}$ tensor we can construct that is linear in curvature is $W_{abcd}k^ak^bk^ck^d$, but this vanishes by virtue of the symmetries of the Weyl tensor. Hence the tensor with the highest powers of $k$ is
\be
A^a{}_{b} = W^a{}_{cbd}k^ck^d \, ,
\ee
and by symmetry of the Weyl tensor we therefore have $A_{ab}k^b = k^b A_{ba}=A^a_a=0$. In a general EFT expansion, we anticipate all scalar local operators to arise suppressed by the cutoff scale. This includes operators that are combinations of $A_{ab}$ contracted with the metric and with itself. Hence following a reasoning identical to that in section \ref{validity}, the highest possible on-shell momenta is determined by (at least) the EFT requirements
\be\label{EFTbounds3}
{\rm Tr}[A^n] \ll M^{4n} \, ,
\ee
for integer $n$. In addition, we must require the more obvious curvature requirement \\
$|W_{abcd}W^{abcd}| \ll M^4$ and related covariant derivative requirements. \\

Since Schwarzschild is time translation invariant, on--shell modes are best characterized in terms of their frequency $\omega \sim i \partial_t$.
For a transverse wave $k_{\mu} = ( - \omega,0,0,\pm \omega r^{1/2} \sin \theta/\sqrt{1-r_g/r} )$ with $k_{\mu}k^{\mu}=0$ the tensor $A_{ab}$ is given by
 \be
 A_{ab} \d x^a \d x^b = \frac{\omega^2 r_g}{2 r^3} \d t^2- \frac{3\omega^2 r_g}{2 r^3(1-r_g/r)^2} \d r^2+ \frac{3\omega^2 r_g}{2 r^3 (1-r_g/r)} r^2 \d \Omega^2 \, ,
 \ee
 and the bounds \eqref{EFTbounds3} amount to the single condition
\be\label{eq:r}
\frac{r_g}{r^3} \frac{ \omega^2}{(1-r_g/r)} \ll M^4 \, .
 \ee

Interestingly, due to the symmetry of the spacetime, for a radial travelling wave $k_{\mu} = (-\omega, \pm \frac{\omega}{(1-r_g/r)} , 0 , 0)$, the situation is then more subtle. We then have
\be
A_{ab} \d x^a \d x^b=-\frac{r_g}{r^3} \omega^2 \( \d t \mp \frac{1}{(1-r_g/r)}  \d r \)^2 \, .
\ee
Since for such a radial mode $A_{ab} \propto k_a k_b$, then the conditions \eqref{EFTbounds3} are automatically satisfied. This does not mean that $\omega$ can be arbitrarily large, but rather that we should look more closely at the higher derivative bounds such as
\be
A^{ab} k^{\mu} \nabla_{\mu} A_{ab} \ll M^8 \ ,
\ee
and
\be
((k^{\mu} \nabla_{\mu})^p A^{ab})( ( k^{\nu} \nabla_{\nu} )^p A_{ab})  \ll M^{8+4p} \, ,
\ee
as well as
\be
(k^{\mu} \nabla_{\mu})^p (W^{abcd} W_{abcd}) \ll M^{4+2p} \, .
\ee
These last two conditions are seen to be strongest in the limit $p \rightarrow \infty$, and amount to $k^{\mu} \nabla_{\mu} \ll M^2$, i.e. to the relation
\be\label{EFTbounds4}
\omega \ll M^2 r \, .
\ee
This is significantly stronger than \eqref{eq:r}, except near the black hole horizon, but applies only for modes with a significant radial component for which $k^{\mu}\nabla_{\mu}=\pm \omega \partial_r+\dots$ picks out the radial dependence of the background geometry.

\subsection{Causality and Time delay}

We now focus on the case of a superluminal radial velocity where $\Delta c_s^2(r)$ given in \eqref{eq:cs} is positive for $r>r_g$, ($2d_9+d_{10}<0$).
As a warm up let us consider the simple analysis of a radial moving trajectory. If we were to simultaneously send an outgoing radial photon and a GW with wavelength $\lambda$ from a distance $r_0>r_g\gg \mpl^{-1}$ just outside the BH horizon, the GW will arrive in advance of the photon at infinity by an amount $\Delta T_{\rm adv}$, whose maximal value is
\ba
\label{eq:DR_BH2}
\Delta T_{\rm adv}&=& \int_{r_0}^\infty  \frac{\d r}{(1-\frac{r_g}{r})} - \int_{r_0}^\infty  \frac{\d r}{c_s(r)(1-\frac{r_g}{r})}\\
&\sim&\int_{r_0}^\infty  \frac{\d r \Delta c_s^2}{2(1-\frac{r_g}{r})}\sim \frac{r_g}{M^2\mpl^2 r_0^4}\,.
\nn
\ea
Remaining within the regime of validity of the EFT and being able to trust this answer requires the bound \eqref{EFTbounds4} to be satisfied at the distance $r_0$, leading to the requirement\ba
\frac{r_g}{r_0^4}\ll M^2\omega^{-1} {\frac{r_g}{r_0^3}} \,.
\ea
This leads to the upper bound
\ba
\label{eq:DR_BH}
\Delta T_{\rm adv}\sim \frac{r_g}{r_0^4}\frac{1}{M^2 \mpl^2}
\ll  \underbrace{\frac{r_g}{r_0}}_{<1}  \underbrace{\frac{1}{\mpl^2 r_0^2}}_{\ll 1} \, \omega^{-1} \ll \omega^{-1}\,,
\ea
where we have used the requirement $W^2\ll \mpl^4$. Consequently the would-be scattering time advance is smaller that the physical wavelength of this mode and is therefore not resolvable. This result remains true regardless of how close to the horizon the initial wave starts out (so long as we only consider waves of frequencies that are within the regime of validity  of the EFT at that point as dictated by \eqref{EFTbounds4}). We see in fact that the effect described by \eqref{eq:DR_BH} is even further suppressed relative to the FLRW effect by a factor $\frac{r_g}{\mpl^2 r_0^3}$ which is consistent with the fact that we are here relying on curvature--cubed (dimension-6) operators rather than curvature--squared (dimension--4) operators.\\

Moving beyond radial trajectories, let us consider the scattering time-delay induced by the black hole background. By symmetry the retarded propagator may be determined in an angular momentum eigenbasis with angular momentum $\ell$ for which the Eisenbud-Wigner time delay  \cite{Wigner:1947zz,Wigner:1955zz,Martin:1976iw} as shown in Appendix~\ref{app:phaseshift} applied to the metric \eqref{Effmetric1} is
\be
\Delta T_\ell  = 2 \int_{r_t}^{\infty} \d r  \left( \frac{1}{Z_t\sqrt{Z_r\left( Z_t^{-1}-Z_{\Omega }^{-1} \frac{b^2}{r^2}  \right)}} - 1 \right) -  2 r_t  \, ,
\ee
where $r_t$ is the turning point where the denominator vanishes $r^2 Z_\Omega=Z_t b^2$.
As discussed in Section~\ref{Shapiro4d} on asymptotically--Schwarzschild spacetimes this is logarithmically divergent in $D=4$, but this need not concern us as the time delay correction to the usual GR Shapiro time--delay is well defined
\be
\Delta T^{\rm EFT}_\ell = 2 \int_{r_t}^{\infty} \d r  \left( \frac{1}{Z_t\sqrt{Z_r\left( Z_t^{-1}-Z_{\Omega }^{-1} \frac{b^2}{r^2}  \right)}} \right)-2 \int_{r_t(0)}^{\infty} \d r  \left( \frac{1}{f(r)\sqrt{1- f(r)\frac{b^2}{r^2}  }} \right) \, ,\nn
\ee
with $f(r) = 1-r_g/r$ and $r_t(0)$ the usual GR turning point. Following the approach given in Appendix~\ref{app:Timedelaycalc} at leading order in the EFT expansion this is
\be
\Delta T^{\rm EFT}_\ell = 2 \int_{r_t(0)}^{\infty} \d r  \left( \frac{1}{f(r)\sqrt{\left( 1- f(r)\frac{b^2}{r^2}  \right)}} \right)  \frac{\d \delta R(r)}{\d r} \, ,
\ee
where the function $\delta R$ is defined in \eqref{eq:deltaR}.
At leading order in a $r_g/b$ expansion is
 \be
 \frac{\d \delta R(r)}{\d r}=  144 r_g (d_{10}+2d_{9}) \frac{(4b^2-r^2)}{M^2 \mpl^2 b^2 r^5} + \dots  \, ,
 \ee
 and yields at leading order in $r_g/b$ a time delay
  \be
 \Delta T^{\rm EFT}_\ell  =  \frac{480}{M^2 \mpl^2}(d_{10}+2d_{9}) \frac{r_g}{b^4} \, .
 \ee
As expected when $d_{10}+2d_{9}<0$, for which radial modes are superluminal, this corresponds to a time advance $ \Delta T^{\rm EFT}_\ell <0$. However we may follow a reasoning similar to that in \eqref{eq:DR_BH} to see that this time advance remains unresolvable within the EFT. Indeed even if we content ourself with the weaker EFT condition \eqref{eq:r} we would  find for $r_g/b \ll 1$
\be\label{resolvable1}
| \Delta T^{\rm EFT}_\ell| \sim  \frac{r_g}{M^2 \mpl^2 b^4} \ll \underbrace{{\sqrt{\frac{r_g}{b}}}}_{<1} \underbrace{\frac{1}{\mpl^2 b^2}}_{\ll 1} \omega^{-1}\ll \omega^{-1} \, ,
\ee
and it is for this reason that causality is not being violated.

\subsection{Failure of the Eikonal/Semiclassical Approximation}

Since the contribution to the scattering phase shift is essentially $\Delta \delta_\ell^{\rm EFT} \sim 2 \omega \Delta T^{\rm EFT}_\ell $ the condition \eqref{resolvable1} is equivalent to the statement that the EFT contribution to the phase shift remains perturbatively small in the regime of validity of the EFT,
\ba
|\Delta \delta^{\rm EFT}_\ell|\ll \sqrt{\frac{r_g}{b} }\frac{1}{\mpl^2 b^2} \ll  1\,.
\ea
Indeed not only is it smaller than unity, even if we were to take the extreme limit $b \rightarrow \mathcal{O}\(r_g\)$, it would still be parametrically suppressed by $(\mpl b)^{-2}$.
Specifically terms of order $(\Delta \delta^{\rm EFT}_\ell)^2$ will be of the same order as other EFT contributions that come in at the order $(\mpl b)^{-4}$. This brings us to the essential point, it is implicit in the eikonal resummation (for example that performed in \cite{Camanho:2014apa}) that the t-channel exchange ladder diagrams dominate over other Feynman diagrams in the perturbative expansion so that it is consistent to resum them into the exponentiated form
\be
\delta_\ell({\text{$t-$channel}})+ \text{higher order ladder diagrams}  \rightarrow \frac{1}{2 i } \(e^{2 i \delta_\ell({\text{$t-$channel}})}-1 \) \, .
\ee
This is a well justified procedure for the GR contributions which give rise to the Shapiro time-delay, and indeed since the Shapiro time-delay is of order $r_g$ times a logarithmic factor,  there is no problem engineering a situation for which $\delta_\ell^g \gg1 $ or equivalently $\Delta T^g \gg \omega^{-1}$ by having $\omega \gg r_g^{-1}$, which is not in conflict with any consistency requirement of pure GR. \\

By contrast, since the EFT contributions to the t-channel exchange are perturbatively small, it is not legitimate to resum them while neglecting other Feynman diagrams that can arise at the same order.
We see that the situation of the low-energy EFT of gravity on a Schwarzschild background is closely analogous to that in FLRW. Performing an eikonal resummation of the contribution $\delta^{\rm EFT}_\ell$ is equivalent to the resummation of the would-be secular terms implicit in \eqref{WKB1} and \eqref{WKB10}. In both cases this resummation is simply not justified at least as an indicator of causal support.
Here, this shows up for us in the fact that the EFT correction to the scattering time delay is not resolvable within the approximation used to calculate the time delay, in analogy with the support of the FLRW lightcone. \\

We stress that our condition for resolvability is not the same as that sometimes required in the literature, namely that the magnitude of the time delay is large in comparison to the naive cutoff, e.g.
\ba
\label{eq:naivebound}
|\Delta T^{\rm EFT}_\ell| \sim \underbrace{\sqrt{\frac{r_g}{b}}}_{< 1}\underbrace{\frac{1}{b M}}_{\ll 1} \underbrace{\frac{1}{b^2 \mpl^2}}_{\ll 1}  M^{-1}\ll M^{-1}\,?
\ea
While we need to ensure that the scattering state/wave remains within the regime of validity of the EFT throughout its trajectory so as to be able to use the low-energy EFT to determine its time-advance, nothing demands that the time-advance itself should be measured within the low-energy EFT. By itself, the bound \eqref{eq:naivebound} is therefore irrelevant. Moreover, $1/M$ is not the cutoff for time measurements within the EFT because the time delay is not a Lorentz invariant quantity. This is why we are careful in sections~\ref{validity} and \ref{validity2} to identify the appropriate cutoff based on locally Lorentz invariant combinations. Resolvability here means whether it can be consistently computed in the semiclassical/eikonal approximation, and at its heart the latter assumes frequencies and wavelengths that are large in comparison to the scales of variation of the background quantities.

\section{QED in Curved Spacetime}
\label{sec:QED}

\subsection{Low--Energy EFT for QED Below the Electron Mass}
It is helpful at this point to compare the above discussion with the classic case of superluminal speeds in a low-energy EFT, that of QED in curved spacetime first pointed out by Drummond and Hathrell \cite{Drummond:1979pp} and extended in \cite{Lafrance:1994in}. The result of \cite{Drummond:1979pp} is particularly clean in that it does not require gravity to be dynamical, i.e. it would be obtained in a decoupling limit $\mpl \rightarrow \infty$ for fixed background curvature. Furthermore on the same background, different polarizations of light can be shown to have low-energy speeds which are both superluminal and subluminal. This gives rise to gravitational birefringence  \cite{Drummond:1979pp} and from higher order operators gravitationally induced dispersion of light \cite{Lafrance:1994in}. The causal implications of this result for the photon have been discussed extensively in the literature \cite{Hollowood:2007kt,Hollowood:2007ku,Hollowood:2008kq,Hollowood:2009qz,Hollowood:2010bd,Hollowood:2010xh,Hollowood:2011yh,Hollowood:2012as}.\\

For our present purposes it is sufficient to note that leading effect from an effective action of the form
\be\label{QED}
{\cal L} = \sqrt{-g} \( - \frac{1}{4} F_{\mu\nu} F^{\mu\nu} + \frac{\alpha}{M^2}  R_{abcd} F^{ab} F^{cd} + \dots \)\,,
\ee
which is the relevant part of the low-energy effective action for QED on a curved spacetime where $M$ is the electron mass. The operator $RFF$ leads to polarization-dependent corrections to the low-energy sound speed. For a transverse-travelling wave (with momentum in the angular direction) on a Schwarzschild background, the low-energy sound speed is of the form \cite{Drummond:1979pp}
\be
\label{eq:csRFF}
c_s^2 =1+ \frac{\beta_{P}}{M^2}\frac{r_g}{r^3}+\mathcal{O}\(\frac{r_g^2}{M^4 r^6}\)\,,
\ee
where $\beta_{P}$ is an order unity polarization dependent constant. Specifically for radially polarized light $\beta_P>0$, while $\beta_P<0$ for angular polarization. Interestingly were \eqref{QED} the exact Lagrangian, we see that the equations for the electromagnetic field remain second order, and this speed would be the group/phase and front velocity. In practice this is not the case due to dispersion from higher order operators in the EFT not included \cite{Lafrance:1994in}. At this level, the low-energy speed \eqref{eq:csRFF} includes no frequency dependence (none of the higher order terms $\mathcal{O}\(\frac{r_g^2}{M^4 r^6}\)$ in \eqref{eq:csRFF} would include any frequency dependence), yet higher--order operators that have not been included in the low-energy EFT \eqref{QED} would affect the speed at high energy and  \eqref{eq:csRFF} can certainly not be the speed of light at arbitrarily high energy.  \\

\subsection{Unresolvable Time Advance} As noted in \cite{Drummond:1979pp} the propagation of a photon can be understood in terms of evolution in an effective metric of the form
 \be\label{Zmetric}
 Z_{\mu\nu} \d x^{\mu} \d x^{\nu} =\(1+ \frac{\beta_{P}}{M^2}\frac{r_g}{r^3}\)\( - f(r) \d t^2 + \frac{1}{f(r)} \d r^2 \)+ r^2 \d^2 \Omega\,,
   \ee
   with the standard $f(r) = (1-\frac{r_g}{r})$. Using the results of appendix \ref{app:phaseshift} for the effective metric \eqref{Zmetric} the naive expression for the Eisenbud-Wigner scattering time delay is
\ba
\Delta T_\ell=2\int_{r_t(\beta_{P})}^\infty\d r \(\frac{1}{f(r)\sqrt{1-\frac{b^2}{r^2}f(r) \(1+ \frac{\beta_{P}}{M^2}\frac{r_g}{r^3}\)}}-1\)-2r_t(\beta_{P})\,,
\ea
where $r_t(\beta_P)$ is the turning point as computed in the EFT with parameter $\beta_P$.
Again in $D=4$ this is logarithmically divergent due to slow Coulomb $1/r$ fall off in four dimensions but what is relevant is the extra time delay relative to scattering in Schwarzschild with $\beta_{P}=0$, i.e.
\be \label{Timeintegral1}
\Delta T^{\rm EFT}_\ell  =2\int_{r_t(\beta_{P})}^\infty\d r \(\frac{1}{f(r)\sqrt{1-\frac{b^2}{r^2}f(r) \(1+ \frac{\beta_{P}}{M^2}\frac{r_g}{r^3}\)}}\)-2 \int_{r_t(0)}^\infty\d r \(\frac{1}{f(r)\sqrt{1-\frac{b^2}{r^2}f(r) }}\) \, ,
\ee
which is a finite expression. As shown in appendix \ref{app:Timedelaycalc}, using \eqref{TD5} to first order in $\beta_{P}$ this is given by
\be
\Delta T^{\rm EFT}_\ell = - \frac{2 \beta_{P} r_g}{M^2} \int_{r_t(0)}^{\infty} \d r \,  \frac{1}{f(r) \sqrt{1-\frac{b^2}{r^2} f(r)}} \frac{\p}{\p r} \left( \frac{ \frac{b^2}{r^5} f(r)^3}{\frac{\p}{\p r} \( \frac{b^2}{r^2} f(r)^3-f(r)^2\)}\right) \, .
\ee
For  large\footnote{We note that the integral on the right hand side of \eqref{Timeintegral1} does become arbitrarily large as $r_t(0)$ approaches $3r_g/2$ which is when $b \rightarrow 3 \sqrt{3} r_g/2$. However this is purely an artifact of the fact that this is the peak of the effective potential \eqref{potentialtortoise}, and is the point at which we may easily transmit across the potential barrier. As such, the boundary conditions used for the solutions to derive \eqref{Timeintegral1} are not appropriate.} impact parameter in comparison to the Schwarzschild radius $b \gg r_g$ we have to leading order in an expansion in $r_g/b$
\ba
\Delta T^{\rm EFT}_\ell=-\frac{2r_g \beta_P}{b^2 M^2}\,,
\ea
in the large $\ell$ limit and to leading order in $r_g/b$. As expected, this is a time advance for $\beta_{P} >0$. \\

At the point of closest approach, the wave is purely transverse and  the bound \eqref{eq:r} applies. Again to leading order in $r_g/b$, we infer that for the EFT to remain valid at the impact parameter, the frequency should satisfy
\ba
\omega \ll M^2 \sqrt{\frac{b}{r_g}}b \,.
\ea
Putting this together the time advance is bounded by
\ba
\label{eq:boundtimeadvanceQED}
|\Delta T^{\rm EFT}_\ell |=2|\beta_P| \(\frac{r_g}{b}\)\frac{1}{b M^2}\ll \sqrt{\frac{r_g}{b}}\omega^{-1} \ll \omega^{-1} \,,
\ea
from which we again conclude that while the low-energy superluminality pointed out by Drummond and Hathrell in the context of the EFT for QED below the electron mass on a curved Schwarzschild background does technically lead to a time-advance $\Delta T_\ell <0$, this advance is not resolvable. In fact since in this example we know the UV completion we can calculate exactly the phase shift, and as shown in \cite{Hollowood:2015elj} in the shockwave limit the contribution to the phase shift is always small ensuring the time-delay is indeed unresolvable. In addition it is shown in \cite{Hollowood:2015elj} that in the limit of high frequencies $\lim_{\omega \rightarrow \infty} \Delta T^{\rm EFT}_\ell(\omega)=0$. This confirms the underlying Lorentz invariant causality of the UV completion, but is secondary to the resolvability criterion in understanding how the low-energy EFT is consistent with causality.

\section{Case of a Resolvable Physical Time Advance}
\label{sec:PX_resolvable}

In the previous sections we have seen how well known examples such as QED in curved spacetime and the EFT of gravity lead to time advances that are nevertheless unresolvable and hence not in tension with causality. To demonstrate that the time-delay analysis is not without content, we consider here an example of the opposite case, an EFT which does lead to a resolvable time advance and can therefore be concluded to be in tension with causality.
For this we need to put ourselves in the situation where the bound \eqref{eq:speed1} is violated in the decoupling limit for which it is sufficient to return to our canonical example of a $P(X)$ scalar field minimally coupled as in  \eqref{EFT1} with the negative parameter $a=-|a|$. \\

In order to parallel the previous Schwarzschild discussion we consider sourcing the scalar field in the manner
\ba\label{sourcecoupling}
\L_{\rm source}=-\frac{\beta}{\mpl}\phi T\,,
\ea
where $T$ is the trace of the stress-energy tensor of all other matter fields present in that spacetime and $\beta$ is a dimensionless coefficient which may be taken parametrically larger than unity. In particular if we consider a situation similar to that of the previous sections, with a Schwarzschild geometry generated by a mass $M_*$  located at $r=0$, with  Schwarzschild  radius $r_g=2M_*/\mpl^2$ so that $T$ itself is a delta function source. Ignoring for now the backreaction of the scalar field onto the geometry (i.e working in the field theory on curved spacetime limit), the background profile $\phi_0(r)$ for the scalar field is determined by the solution of\footnote{We stress it is not important that there is a global solution of this equation for all $r$, as we are working with a truncated EFT and the solution can only be trusted in the regime given typically by \eqref{safeeft}.}
\ba
\phi_0'(r)\left[1+|a|f(r)\frac{\phi_0'(r)^2}{M^4}\right]=-\frac{\beta r_g \mpl}{2 r^2 f(r)}\,,
\ea
so that in the weak field regime,
\ba
\phi_0'= -\frac{\beta r_g \mpl}{2 r^2 f(r)} \(1+\mathcal{O}\( \beta^2 \frac{r_g^2 \mpl^2}{r^4 M^4}\) \)\,.
\ea
To remain within the safe region of the EFT we require $|X|\ll M^4$ and so
\be\label{safeeft}
\beta^2 \frac{r_g^2 \mpl^2}{f(r)r^4 M^4} \ll 1\, .
\ee
Considering fluctuations about this background, $\phi=\phi_0(r)+\delta \phi(t,r,\varphi)$, the angular speed of $\delta \phi$-waves remain luminal while the radial speed is not only superluminal for $a<0$,
\ba
c_r^2=1-a\frac{\beta^2 \mpl^2 r_g^2}{2f(r)r^4 M^4}\,,
\ea
the departure from luminality is enhanced by a factor of $(\mpl^2/M^2)(r_g/r)$ as compared with that in QED \eqref{eq:csRFF} (and by a factor of $(\mpl^2/M^2)(\mpl^2 r_g^2)(r/r_g)\ggg 1$ as compared to the EFT of gravity \eqref{eq:cs}). More precisely the effective metric for the fluctuations of the scalar field $\delta \phi$ to leading order in the EFT expansion is of the form
\be
Z_{\mu\nu} \d x^{\mu} \d x^{\nu} = \(1+|a|\frac{\beta^2 \mpl^2 r_g^2}{2f(r)r^4 M^4}\)\( - f(r) \d t^2 +r^2 \d^2 \Omega\)+ \frac{1}{f(r)} \d r^2  \, .
\ee
Once again following the approach of appendix \ref{app:Timedelaycalc}, the leading EFT correction to the time delay relative to Schwarzschild is
\be
\Delta T^{{\rm EFT}, P(X)}  = 2 \int_{r_t(0)}^{\infty} \d r \frac{1}{f(r) \sqrt{1-\frac{b^2}{r^2}f(r)}} \frac{\d \delta R(r)}{\d r } \, ,
\ee
where here
\be
\frac{\d \delta R(r)}{\d r } = \frac{a \beta^2 \mpl^2 r_g^2 (2 b^4 (3r-5r_g)(r-r_g)^2-4 r^6 r_g+b^2 r^3(-2 r^2+8r r_g-5 r_g^2))}{2 M^4 r^3 \(b^2 (2r-5r_g)(r-r_g)+2r^3 r_g\)^2} \, .
\ee
This simplifies considerably in the limit of large impact parameter $b \gg r_g$ to be
\be
\frac{\d \delta R(r)}{\d r } = \frac{a \beta^2 \mpl^2 r_g^2 (3 b^2-r^2)}{4 b^2 M^4 r^4} \, .
\ee
which gives to leading order in $r_g/b$ the time delay correction
\ba
\Delta T_\ell^{{\rm EFT}, P(X)} \sim 2 \int_{b}^{\infty} \d r \frac{1}{ \sqrt{1-\frac{b^2}{r^2}}} \frac{a \beta^2 \mpl^2 r_g^2 (3 b^2-r^2)}{4 b^2 M^4 r^4} =\frac{a \beta^2 \mpl^2 r_g^2 \pi }{8 b^3 M^4 }   \, ,
\ea
which as expected is a time advance for $a<0$.

Now turning back to establishing the regime of validity of the EFT, since the angular speed is luminal, in this case the time delay is maximal at small angular momentum. Unlike the previous discussion in section \ref{validity2} already for a radial wave of the form  \mbox{$k_\mu=(-\omega,\pm \omega f(r)^{-1}(1+a \frac{\phi_0'^2 f}{M^4}),0,0)$}, the bounds \eqref{EFTbounds3} are non-trivial when the curvature of the effective metric $Z_{\mu\nu}$ is used and lead to the requirement
\ba
b^7\gg \frac{\beta^2 \mpl^2 r_g^3}{M^8} \omega^2\,.
\ea
Using this bound in the time advance, we get a bound in terms of the inverse frequency
\ba
\left|\Delta T_\ell^{{\rm EFT}, P(X)} \right| \sim \frac{\beta^2 \mpl^2 r_g^2 }{b^3M^4}
\ll   |\beta| \sqrt{\mpl^2 r_g b} \, \omega^{-1}\,.
\ea
Remarkably we are now here in a situation where this bound can be satisfied with a resolvable time advance $|\Delta T_\ell^{{\rm EFT}, P(X)}|  \gg \omega^{-1}$ so long as $\mpl^2 r_g b \sim M_\star b \gg 1$ which is extremely easy to satisfy. Take the sun for instance at its Schwarzschild radius $b\sim r_g$ would lead to a factor $ \sqrt{\mpl^2 r_g b} \sim M_\star/\mpl \sim 10^{38}$. \\

However \eqref{EFTbounds3} is only one possible EFT condition and we should also check that this effect survives for example \eqref{EFTbounds4} or \eqref{eq:r}. Applying the more conservative criterion \eqref{EFTbounds4} to be safe we would be led to
\ba \label{bound20}
\left|\Delta T_\ell^{{\rm EFT}, P(X)} \right| \sim \frac{\beta^2 \mpl^2 r_g^2 }{b^3 M^4}
\ll   \beta^2 \frac{\mpl^2}{M^2} \frac{r_g^2}{b^2} \, \omega^{-1}\,.
\ea
Again we can engineer a resolvable time advance with $|\beta| \mpl r_g \gg M b$ and \eqref{safeeft} can still be satisfied for $b \gg M^{-1}$. The simplest way to achieve this is to make $| \beta| \gg 1$. Indeed, since the theory defined by \eqref{EFT1} has a well defined decoupling limit as $\mpl \rightarrow \infty$, the time advance has no relationship to gravity per se. The coefficient $\beta$ is the source coupling \eqref{sourcecoupling} may equally well have been defined as $\beta  = \tilde \beta \mpl/M$ so that the coupling  \eqref{sourcecoupling} survives in the limit $\mpl \rightarrow \infty$ for fixed source mass. Then the bound \eqref{bound20} may equally well be written as
\be
\left|\Delta T_\ell^{{\rm EFT}, P(X)} \right| \ll   |\tilde \beta| \frac{M_*^2}{M^4 b^2} \, \omega^{-1}\,.
\ee
In terms of scattering amplitudes, this can be related to a phase shift of order
\ba
\left| \Delta \delta_\ell^{{\rm EFT}, P(X)}\right|\ll  |\tilde \beta|\frac{M_*^2}{M^4 b^2}  \,,
\ea
which may easily be engineered to be $\left|\Delta \delta_\ell^{{\rm EFT}, P(X)}\right| \gtrsim 1$, by taking $M_*^2/M^2 \gg M^2 b^2$ while maintaining $b\gg M^{-1}$ to ensure \eqref{safeeft}.  For such a phase shift, the implicit summation that goes into the semiclassical/eikonal approximation is justified as the ladder diagrams dominate over other contributions at each order in loops. This situation is orthogonal to that of the EFT of gravity and that of QED below the electron mass and as discussed in section \ref{suppression} is due to the lack of suppression of the speed correction, a consequence of the fact that \eqref{eq:speed1} is violated.

\section{Discussion}
\label{sec:discussion}

To summarize, when dealing with a gravitational effective theory, superluminal low-energy speeds (with respect to the metric out of which it is constructed) is not only possible, but is sometimes demanded by underlying causality and analyticity criteria. Indeed this has been well known since the work of \cite{Drummond:1979pp,Lafrance:1994in} for QED in curved spacetime \cite{Hollowood:2007kt,Hollowood:2007ku,Hollowood:2008kq,Hollowood:2009qz,Hollowood:2010bd,Hollowood:2010xh,Hollowood:2011yh,Hollowood:2012as}, and more recently noted in \cite{deRham:2019ctd,deRham:2020ejn} for GWs. Relativistic causality nevertheless remains intact because the causal support of the retarded propagator vanishes outside of the lightcone of the metric, defined in the field frame with a well--defined decoupling limit. This apparent contradiction is resolved in the examples of FLRW and Schwarzschild we have discussed by identifying the regime of validity of the effective theory, and asking -- given the low-energy form of the retarded propagator -- whether it is possible to influence events outside of the metric lightcone? When the propagator is computed perturbatively, this can only happen when there are secular terms in the perturbative expansion which need to be resummed, and it is this resummation that extends the support of the lightcone. This is exactly what happens in known pathological cases where relativistic causality is violated \cite{Aharonov:1969vu,Adams:2006sv}.  Here we have shown that in the case of EFT corrections discussed in \cite{deRham:2019ctd,deRham:2020ejn}, the condition for the validity of the EFT automatically precludes any secular behaviour and hence relativistic causality is left intact, despite explicitly superluminal low-energy speeds.\\

With this in mind, faced with a given gravitational effective theory, how should we apply causality requirements in the absence of clean S-matrix requirements? As stated in the introduction many works in the literature simply demand that around a given background, the speed of propagation of all modes is (sub)luminal relative to the metric out of which the theory is constructed. This is a demonstrably false criterion as it is not invariant under field redefinitions as the example in section \ref{fieldframeexample} clearly demonstrates. As noted above, the issue about field frame dependence can be partially resolved by working in a frame in which it is possible to consistently take a decoupling limit $\mpl \rightarrow \infty$ for which gravitational effects decouple (see section~\ref{sec:DL} for clarifications on what is meant by a decoupling limit).
It is then consistent to demand that the resulting Minkowski spacetime field theory respects all of the standard causality/analyticity/positivity requirements. In particular this leads to the condition \eqref{eq:speed1}. Once this is done, this implies that all effects that lead to mild superluminalities are in this frame $\mpl$ suppressed effects. \\

The next question is how do we impose causality away from the decoupling limit, at finite $\mpl$, or in situations in which there is no clean decoupling limit? Once again, simply demanding $c_s  \le 1$ for all perturbations around a given background is an incorrect criterion, as the known examples illustrate \cite{Drummond:1979pp,Lafrance:1994in,deRham:2019ctd,deRham:2020ejn}. Ideally we may just appeal to the full UV theory to demonstrate that the exact retarded propagator is causal \cite{Hollowood:2007kt,Hollowood:2007ku,Hollowood:2008kq,Hollowood:2009qz,Hollowood:2010bd,Hollowood:2010xh,Hollowood:2011yh,Hollowood:2012as}, however we rarely have that luxury, and furthermore in time-dependent spacetimes such as FLRW such analyticity methods are not applicable. We can however cleanly answer this equation within the low-energy effective theory itself provided that we appropriately identify its regime of validity. Indeed causality resolution should entirely lie within the purview of the low-energy EFT as by construction it gives the large distance macroscopic description of the theory, and this is where any causality violation will become apparent. If the superluminal speed were physical, it would lead to a macroscopic consequence, specifically an enlarged support for the retarded propagator. Since at zeroth order in the EFT expansion we assume that the retarded propagator has the support implied by the usual metric lightcone, and since we can structure the calculation of the corrected retarded propagator as a perturbative expansion around this, the telltale sign for a modified causal structure is secular behaviour in the perturbative expansion that needs to be resummed. We have shown explicitly that in the cases considered in \cite{deRham:2019ctd,deRham:2020ejn} this secular behaviour is absent provided we restrict ourselves to effects which can be consistently calculated in the regime of validity of the EFT in question. Concretely in FLRW validity of the EFT imposes a cutoff in on-shell momenta of the form
\be
\frac{| \vec k|}{a(t)}  \ll \Lambda_{c} = \frac{M^2}{\sqrt{- \dot H}} \, .
\ee
This cutoff is parametrically higher than the invariant cutoff $M$ due to the underlying assumption of Lorentz invariance of the UV completion\footnote{Since causality is clearly very different for fundamentally Lorentz violating theories, we do not discuss them here.}. \\

The entire discussion for cosmological spacetimes parallels exactly the more straightforward case of scattering in asymptotically flat/Schwarzschild geometries. There we may use the positivity of the Eisenbud-Wigner scattering time delay as a clean criterion for causality. However any computed time advance from EFT corrections can only be interpreted as a macroscopic causality violation if it is resolvable meaning if
$\Delta T^{\rm EFT}_{\ell} \gtrsim \omega^{-1}$. In all the known cases of time advances arising from consistent UV completions, the time advance is not resolvable and its associated contribution to the scattering phase shift is less than unity $|\Delta \delta^{\rm EFT}_{\ell}| \ll 1$ and $\Delta T^{\rm EFT}_{\ell} \ll \omega^{-1}$. Hence the eikonal/semiclassical resummation, which is the parallel of the secular resummation of the retarded propagator, is not justified as an indicator of causal properties.
Explicit UV completions show that the high energy behaviour of the corrections to the phase shift are very different, and generically we expect $\lim_{\omega \rightarrow \infty} \Delta T^{\rm EFT}(\omega) \rightarrow 0$, equivalent to the expectation that in an underlying Lorentz invariant theory $\lim_{\omega \rightarrow \infty} c_s(\omega) \rightarrow 1$. We stress again however that we are not using these expectations to resolve the naive causality issue with the low-energy EFT. \\

In using causality to constraint EFTs, in the end the issue is not just about the signs of Wilson coefficients in an effective action, but also crucially about their sizes. If a given operator induces a large superluminal effect that is both (a) nonzero in the decoupling limit, and (b) leads to secular behaviour within the regime of validity, then we can safely conclude that this would violate the traditional requirements of relativistic causality as in the case of the `wrong sign' $P(X)$ model. However, if a given operator leads to a {\it small} superluminal speed which is insufficient to give rise to any secular growth in the regime of validity of the EFT, then we cannot conclude any conflict with relativistic causality. Although we have not emphasized this here, there is a strong interplay between these requirements and the application of positivity bounds in a gravitational setting as discussed in \cite{deRham:2019ctd}. This weaker requirement of `signs' of EFT coefficients likely connects with weaker requirements for positivity bounds in the presence of a massless graviton as discussed in \cite{Alberte:2020jsk,AlberteToAppear}. \\

In view of the previous discussion, one may be tempted to
hastily conclude that superluminalities in a low-energy EFT could potentially be allowed so long as they are bounded to be highly suppressed, along the lines of $c_s^2<1+\mathcal{O}(\mpl^{-2})$ and such considerations are hence irrelevant for realistic observations. Such a conclusion would be incorrect. As emphasised in section~\ref{sec:decouplingLim}, great care ought to be given to dealing with superluminalities in the gravitational setup and the bound \eqref{bound4} only makes sense if and in a frame where the decoupling limit $\mpl\to \infty$ can be taken appropriately. What this implies in practise depends on context and on the precise frame in which the bounds are determined, but in many EFTs for inflation and dark energy this often implies that the correct bound to be imposed on the speed of gravitational waves following from causality is  in fact in the regime in which they are strictly superluminal rather than subluminal. Biasing data with a subluminal speed of gravitational wave prior can often turn out to be in direct tension with causality.  

\bigskip
\noindent{\textbf{Acknowledgments:}}
 The work of AJT and CdR is supported by an STFC grant ST/P000762/1. CdR thanks the Royal Society for support at ICL through a Wolfson Research Merit Award. CdR is supported by the European Union's Horizon 2020 Research Council grant 724659 MassiveCosmo ERC--2016--COG and by a Simons Foundation award ID 555326 under the Simons Foundation's Origins of the Universe initiative, `\textit{Cosmology Beyond Einstein's Theory}'. AJT thanks the Royal Society for support at ICL through a Wolfson Research Merit Award.

\appendix

\section{Low-energy EFT for Gravity}
\label{app:EFT}

\subsection{Graviton Dispersion Relation}

The leading modification to the dispersion relation of a gravitational wave in a curved background can be determined by first inspecting the effect in Minkowski spacetime. That is because diffeomorphism invariance may be used to covariantize the Minkowski answer and the ambiguities in this procedure turn out to be subleading corrections. With this in mind let us first determine the form of the graviton propagator or two point function in Minkowski spacetime. At the linear level this may be described by a Lorentz tensor $h_{\mu\nu}$ which transforms under linear diffeomorphisms as $h_{\mu\nu} \rightarrow h_{\mu\nu} +\partial_{\mu} \xi_{\nu}+ \partial_{\mu} \xi_{\nu}$. Although nonlinearly we cannot construct local gauge invariants for gravity, at the linear level we can by means of introducing a conserved external source $T_{\mu\nu}$ for which $\partial_{\mu} T^{\mu\nu}=0$, and considering the `$TT$ amplitude' that describes the interaction between two sources (equivalently the free field connected generating function)
\be\label{TTdef}
\Delta S_{TT} =\frac{i}{2 \mpl^2}\int \d^4 x \int \d^4 y \,  T^{\mu\nu}(x) \langle 0 | \hat T h_{\mu\nu}(x)  h_{\alpha \beta}(y) | 0 \rangle T^{\alpha\beta}(y) \, .
\ee
Following standard arguments, causality, unitarity and Lorentz invariance fix the form of this amplitude to be
\be
\Delta S_{TT}  =  \int \frac{\d^4 k}{(2\pi)^4} T^{\mu\nu}(-k) G_{\mu\nu\alpha \beta}(k) T^{\alpha \beta}(k) \, ,
\ee
where formally
\be\label{propagator}
G_{\mu\nu\alpha \beta}(k)=\frac{1}{2} \frac{Z}{\mpl^2}\frac{P^{2,0}_{\mu\nu\alpha \beta}}{k^2}
+ \frac{1}{\mpl^4} \int_0^{\infty} \d \mu \rho_2(\mu) \frac{P^{2}_{\mu\nu\alpha \beta} }{\mu+k^2}+ \frac{1}{\mpl^4}\int_0^{\infty} \d \mu \rho_0(\mu) \frac{P^{0}_{\mu\nu\alpha \beta}}{\mu+k^2}\,,
\ee
and where the polarization tensors are given by
\ba
&& P^{2,0}_{\mu\nu\alpha \beta} =\frac{1}{2}(\eta_{\mu\alpha} \eta_{\nu \beta}+ \eta_{\nu\alpha} \eta_{\mu \beta}-\eta_{\mu\nu} \eta_{\alpha \beta} ) \, ,\\
&& P^{2}_{\mu\nu\alpha \beta} =\frac{1}{2}(\eta_{\mu\alpha} \eta_{\nu \beta}+ \eta_{\nu\alpha} \eta_{\mu \beta}-\frac{2}{3}\eta_{\mu\nu} \eta_{\alpha \beta}) \, ,\\
&& P^{0}_{\mu\nu\alpha \beta} =\eta_{\mu\nu} \eta_{\alpha \beta} \, .
\ea
In writing this expression we assume that this is the true propagator and so is independent of UV cutoffs or RG sliding scales used in computing loops. The first term in \eqref{propagator} is the usual massless graviton pole which must arise by diffeomorphism invariance uncorrected other than by a wavefunction renormalization\footnote{As famously noted in \cite{PhysRev.125.397} we can have $Z=0$ and hence mass with gauge invariance. This is realized explicitly in the DGP model where the massless graviton mode is non-normalizable \cite{Dvali:2000hr}.}. The second and third terms come from the exchange of intermediate massive states of spin-2 and spin-0 which may arise from either tree level exchange or loop corrections. Unitarity here comes from the strict positivity requirement that
\be
\rho_2(\mu)>0 \, \quad \text{ and } \quad \rho_0(\mu) >0 \, .
\ee

\subsection{Renormalization group without renormalization}

In practice however \eqref{propagator} is only valid if the integrals $ \int_{\mu_0}^{\infty} \d \mu \rho_2(\mu) \mu^{-1}$ and $ \int_{\mu_0}^{\infty} \d \mu \rho_0(\mu) \mu^{-1}$ converge for some finite $\mu_0>0$, i.e. provided they converge in the UV. In general they do not and their divergence is directly related to the renormalization of the curvature squared terms in the effective action. As such on dimensional grounds we expect them to be logarithmically divergent, which is borne out by explicit calculation at one-loop \cite{deRham:2019ctd}.

To deal with this we perform one subtraction defined at an arbitrary scale $\mu_0$ to give
\ba\label{propagator2}
&& G_{\mu\nu\alpha \beta}(k)=\frac{1}{2} \frac{Z}{\mpl^2}   \frac{P^{2,0}_{\mu\nu\alpha \beta}}{k^2} +\frac{C_{2}(\mu_0)}{\mpl^4}P^{2}_{\mu\nu\alpha \beta}+\frac{C_{0}(\mu_0)}{\mpl^4}P^{0}_{\mu\nu\alpha \beta}   \\
&& + \frac{(\mu_0-k^2)}{\mpl^4} \int_0^{\infty} \d \mu \rho_2(\mu) \frac{P^{2}_{\mu\nu\alpha \beta}}{(\mu+\mu_0)(\mu+k^2)}+ \frac{(\mu_0-k^2)}{\mpl^4} \int_0^{\infty} \d \mu \rho_0(\mu) \frac{P^{0}_{\mu\nu\alpha \beta}}{(\mu+\mu_0)(\mu+k^2)} \, . \nn
\ea
Since the propagator cannot depend on the arbitrary subtraction scale $\mu_0$ we obtain the dispersion relation analogue of the renormalization group equations
\be\label{RGeq}
\mu_0 \frac{\d }{\d \mu_0} C_S(\mu_0) =-\int_0^{\infty} \d \mu \rho_S(\mu) \frac{\mu_0}{(\mu+\mu_0)^2} \, ,
\ee
the right hand side being finite if our assumption about the overall number of subtractions were correct. We stress though that $\mu_0$ should not be confused with any UV cutoff or sliding scale used in computing loops, it is however clear that it plays a similar role. The integration constant that arises in the solution of the equation are the undetermined subtraction constants. It is natural to define the IR values of these constants at $\mu_0=M_{\rm IR}^2$, and the UV at some high energy scale $\mu_0=M_{\rm UV}^2$. Integrating the RG equation we have
\be
C_S^{\rm IR}= C_S^{\rm UV} + \int_0^{\infty} \d \mu \rho_S(\mu) \frac{(M_{\rm UV}^2-M_{\rm IR}^2)}{(\mu+M_{\rm IR}^2)(\mu+M_{\rm UV}^2)} \, ,
\ee
and so we clearly have by unitarity
\be \label{flow}
C_S^{\rm IR}>C_S^{\rm UV} \, .
\ee
Thus the dispersion relation demands positivity of the flow from the UV to the IR. It does not however guarantee that $C_S^{\rm IR}>0$.

\subsection{Low energy effective field theory}

Let us now make an assumption similar to that described in section~\ref{sec:refractive} that the dominant contribution to the spectral densities comes from energies for which $\mu\ge M^2$, where $M$ is viewed as the cutoff of the low-energy effective theory. In other words we assume that there is a weakly coupled low-energy effective theory for which the loop contributions from light fields are small relative to the effects from heavy fields whose masses satisfy $M_I\ge M$. This is quite natural here for situations in which the number of heavy fields with masses greater than $M$ is much larger than the number of light fields, precisely because each field at one-loop contributions logarithmically to $C_S$.

In this case the propagator may be split up as
\be\label{decomposition}
G_{\mu\nu\alpha \beta}(k)=G^{\rm EFT}_{\mu\nu\alpha \beta}(k)+G^{\rm IR}_{\mu\nu\alpha \beta}(k) \, ,
\ee
where the IR part comes exclusively from loops of light fields of masses smaller than $M$, and is thus parametrically suppressed if the low-energy effective theory is weakly coupled
\ba
G^{\rm IR}_{\mu\nu\alpha \beta}(k) &=&  \frac{(\mu_0-k^2)}{\mpl^4} \int_0^{M^2} \d \mu \rho_2(\mu) \frac{P^{2}_{\mu\nu\alpha \beta}}{(\mu+\mu_0)(\mu+k^2)}\\
&+& \frac{(\mu_0-k^2)}{\mpl^4} \int_0^{M^2} \d \mu \rho_0(\mu) \frac{P^{0}_{\mu\nu\alpha \beta}}{(\mu+\mu_0)(\mu+k^2)} \, ,\nn
 \ea
and the remaining part is that which will essentially be described by the tree level low-energy effective theory
\ba\label{propagator3}
&& G^{\rm EFT}_{\mu\nu\alpha \beta}(k) = \frac{1}{2} \frac{Z}{\mpl^2}   \frac{P^{2,0}_{\mu\nu\alpha \beta}}{k^2} +\frac{C_{2}(\mu_0)}{\mpl^4}P^{2}_{\mu\nu\alpha \beta}+\frac{C_{0}(\mu_0)}{\mpl^4}P^{0}_{\mu\nu\alpha \beta}   \\
&& + \frac{(\mu_0-k^2)}{\mpl^4} \int_{M^2}^{\infty} \d \mu \rho_2(\mu) \frac{P^{2}_{\mu\nu\alpha \beta}}{(\mu+\mu_0)(\mu+k^2)}+ \frac{(\mu_0-k^2)}{\mpl^4} \int_{M^2}^{\infty} \d \mu \rho_0(\mu) \frac{P^{0}_{\mu\nu\alpha \beta}}{(\mu+\mu_0)(\mu+k^2)} \, . \nn
\ea
The IR part $G^{\rm IR}_{\mu\nu\alpha \beta}(k)$ is generically non-local since it includes loops of light fields, in particular those of the graviton itself. For example if $\rho_{0,2}(\mu)$ are approximately constant over the range $0\le \mu < M^2$ then we have approximately
\be
\hspace{-0.3cm} G^{\rm IR}_{\mu\nu\alpha \beta}(k) \approx  \frac{1}{\mpl^4}  \rho_2(0) P^{2}_{\mu\nu\alpha \beta}\ln\(\frac{\mu_0(M^2+k^2)}{k^2(M^2+\mu_0)} \)+ \frac{1}{\mpl^4}  \rho_0(0) P^{0}_{\mu\nu\alpha \beta} \ln\(\frac{\mu_0(M^2+k^2)}{k^2(M^2+\mu_0)} \)\, .
 \ee
By contrast $ G^{\rm EFT}_{\mu\nu\alpha \beta}(k) $ is  local when viewed at energies $|k^2| \ll M^2$. In other words we may perform a standard EFT expansion in the form
\ba\label{propagator4}
&& G^{\rm EFT}_{\mu\nu\alpha \beta}(k) \approx \frac{1}{2} \frac{1}{\mpl^2} Z  \frac{P^{2,0}_{\mu\nu\alpha \beta}}{k^2} +\frac{C_{2}(\mu_0)}{\mpl^4}P^{2}_{\mu\nu\alpha \beta}+\frac{C_{0}(\mu_0)}{\mpl^4}P^{0}_{\mu\nu\alpha \beta}   \\
&& +\sum_{n=1}^{\infty} \frac{(\mu_0-k^2)^n}{\mpl^4} \int_{M^2}^{\infty} \d \mu \rho_2(\mu) \frac{P^{2}_{\mu\nu\alpha \beta}}{(\mu+\mu_0)^{n+2}}+ \sum_{n=1}^{\infty} \frac{(\mu_0-k^2)^n}{\mpl^4} \int_{M^2}^{\infty} \d \mu \rho_2(\mu) \frac{P^{0}_{\mu\nu\alpha \beta}}{(\mu+\mu_0)^{n+2}}\, . \nn
\ea
Furthermore taking $\mu_0=M_{\rm IR}^2 \approx 0$, then in the region $M_{\rm IR}^2 \ll |k^2| \ll M^2$ we may approximate this as
\ba\label{propagator5}
&& G^{\rm EFT}_{\mu\nu\alpha \beta}(k) \approx \frac{1}{2} \frac{1}{\mpl^2} Z  \frac{P^{2,0}_{\mu\nu\alpha \beta}}{k^2} +\frac{C_{2}^{\rm IR}}{\mpl^4}P^{2}_{\mu\nu\alpha \beta}+\frac{C_{0}^{\rm IR}}{\mpl^4}P^{0}_{\mu\nu\alpha \beta}   \\
&& +\sum_{n=1}^{\infty} \frac{(-k^2)^{n}}{\mpl^4} \int_{M^2}^{\infty} \d \mu \rho_2(\mu) \frac{P^{2}_{\mu\nu\alpha \beta}}{\mu^{n+2}}+ \sum_{n=1}^{\infty} \frac{(-k^2)^n}{\mpl^4} \int_{M^2}^{\infty} \d \mu \rho_2(\mu) \frac{P^{0}_{\mu\nu\alpha \beta}}{\mu^{n+2}}\, . \nn
\ea
This is the standard form of the a tree level effective theory description of the propagator as a local derivative expansion, and is the direction analogue of \eqref{Taylor1}.

\subsection{1PI and Wilson Effective action}

In performing the decomposition \eqref{decomposition}, we are implicitly assuming that the light loops are computed through a unitarity cut method, consistent with the dispersion relation. In short, rather than computing the loop process through standard means, we compute its imaginary part, and then infer its remaining contribution through its dispersion relation. With this proviso, we then recognize that $ G^{\rm EFT}_{\mu\nu\alpha \beta}(k)$ will be the two point function computed from the Wilsonian effective action valid below the scales $|k^2|< M^2$, and $G_{\mu\nu\alpha \beta}(k)=G^{\rm EFT}_{\mu\nu\alpha \beta}(k)+G^{\rm IR}_{\mu\nu\alpha \beta}(k)$ will be the result of the 1PI effective action. As always this split is arbitrary, here depending on the subtraction scale $\mu_0$ which hence we take as some IR scale $\mu_0 = M_{\rm IR}^2$. \\

Assuming we only compute loops of heavy fields (not the graviton itself), then the 1PI effective action and Wilsonian effective actions will be diffeomorphism invariant. We may thus write a local covariant action which can reproduce the propagator $G^{\rm EFT}_{\mu\nu\alpha \beta}(k) $ which is found to be
\be\label{eq:covariantaction}
{\cal L}^{\rm EFT} = \sqrt{-g} \left( \frac{\mpl^2}{2} R +C_0^{\rm IR} R^2   + C_2^{\rm IR} (R_{\mu\nu} R^{\mu \nu}- \frac{1}{3} R^2 ) \right) + \text{higher derivative terms} \, .
\ee
The higher derivatives terms are relevant for BH solutions, but these cannot be inferred from our above argument and must be computed explicitly as done in \cite{deRham:2019ctd,deRham:2020ejn}. However they do have the virtue of being prescription independent. We can also add to this action a Gauss-Bonnet term which cannot be inferred from our calculation. Indeed up to a Gauss-Bonnet term \eqref{eq:covariantaction} may equivalently be written as
\be\label{eq:covariantaction}
{\cal L}^{\rm EFT} = \sqrt{-g} \left( \frac{\mpl^2}{2} R + C_{R^2}^{\rm IR} R^2   +C_{W^2}^{\rm IR} W_{\mu\nu\alpha\rho}^2 + C_{\rm GB} {\rm GB}  \right) + \text{higher derivative terms} \, ,
\ee
where
\be
C_{W^2}=\frac{1}{2} C_2 \, \quad C_{R^2}=C_0 \, .
\ee
The virtue of writing this covariantly is that assuming that matter remains minimally coupled to this metric then we may infer the effect of these additional corrections on other backgrounds such as FLRW. The positivity of the flow \eqref{flow} then implies
\be
C_{W^2}^{\rm IR} > C_{W^2}^{\rm UV} \,,  \quad  C_{R^2}^{\rm IR} > C_{R^2}^{\rm UV}  \, .
\ee

The covariant form of the 1PI effective action being non-local is more complicated (see \cite{Barvinsky:1985an,Barvinsky:1995jv,Barvinsky:1994ic,Avramidi:1990je,Avramidi:1990ap,Barvinsky:1993en,Barvinsky:1994hw,Vilkovisky:2007ny,Codello:2012kq}), but it is sufficient to note that the following covariant expression reproduces the desired $TT$ amplitude and is consistent with the Wilsonian effective action
\ba\label{1PI}
&& {\cal L}^{\rm 1PI}=\sqrt{-g} \left( \frac{\mpl^2}{2} R +C_{2}(\mu_0) (R_{\mu\nu} R^{\mu \nu}- \frac{1}{3} R^2 )+C_{0}(\mu_0)R^2+ C_{\rm GB}(\mu_0)R^{\mu\nu\rho\sigma}R_{\mu\nu\rho\sigma}^*\right.   \nn \\
&& \left.  + R^{\mu\nu}  \int_0^{\infty} \d \mu \rho_2(\mu) \frac{(\mu_0+\Box)}{(\mu+\mu_0)(\mu-\Box)} (R_{\mu\nu}- \frac{1}{3} g_{\mu\nu} R)  + R  \int_0^{\infty} \d \mu \rho_0(\mu) \frac{(\mu_0+\Box)}{(\mu+\mu_0)(\mu-\Box)} R\, \right.  \nn \\
&& \left.   + R^{\mu\nu\rho\sigma}  \int_0^{\infty} \d \mu \rho_{\rm GB}(\mu) \frac{(\mu_0+\Box)}{(\mu+\mu_0)(\mu-\Box)}  R_{\mu\nu\rho\sigma}^* \right) + \dots \, .
\ea
The last term is a non-local extension of the Gauss-Bonnet whose coefficient cannot be inferred from the arguments made so far. In particular we cannot assume that $\rho_{\rm GB}(\mu) >0$. As a special case, a non-local action of this form is used for example in \cite{Donoghue:2014yha} for the specific form of $\rho_2(\mu)$ and $\rho_1(\mu)$ that arises from one-loop integrals of massive and massless states. We stress again that despite appearances \eqref{1PI} is independent of the arbitrary subtraction scale $\mu_0$.

\section{Semiclassical Phase Shift and Time delay}\label{app:phaseshift}

\subsection{Langer approach}
The scattering phase shift $\delta_\ell$ in the semiclassical (WKB) approximation for scattering in a spherically symmetric background is easily computed and we sketch the essential result here. We will assume that the equation of motion of the propagating degrees of freedom may be put in the form of a scalar field living on an effective metric $Z_{\mu\nu}$ as is the case in all the discussed examples. Although generically this equation will have an effective mass, this mass term makes a negligible contribution to the scattering phase shift for high frequencies, consequently it is sufficient to consider a massless scalar for the purposes of our discussion.
To accommodate the behaviour of modes with $\ell$ small we follow the approach of Langer \cite{Langer:1937qr}. The fluctuations of the effective scalar in a $D$--dimensional spherically symmetric background expressed in the form
\be
\label{eq:ZZ}
Z\mn \d x^\mu \d x^\mu=- Z_t\, \d t^2+Z_r^{-1}\d r^2+  r^2\, Z_\Omega\, \d \Omega_{D-2}^2\,,
\ee
can be expressed in terms of $D-2$ dimensional generalization of spherical harmonics, i.e. eigenstates of $\nabla_{D-2}^2 = - \ell(\ell+(D-3))$ and satisfies a wave equation for a given $\ell$
\be\label{eq:wave}
{\cal Z}_t\omega^2 \phi_\ell + \frac{1}{r^{D-2}} \frac{\partial}{\partial r}\left(  r^{D-2} {\cal Z}_r \frac{\partial}{\partial r} \phi_\ell \right) - {\cal Z}_{\Omega }\frac{\ell(\ell+D-3)}{r^2} \phi_\ell = 0 \, ,
\ee
where ${\cal Z}_t = \sqrt{Z_t Z_r^{-1} } Z_\Omega^{(D-2)/2} Z_t^{-1}$, ${\cal Z}_r = \sqrt{Z_t Z_r^{-1} } Z_\Omega^{(D-2)/2} Z_r $ and ${\cal Z}_{\Omega }\ = \sqrt{Z_t Z_r^{-1}}Z_\Omega^{(D-2)/2} Z_{\Omega}^{-1}  $.  \\

We will initially assume that  $Z_{\mu\nu}$ has no singularity, and no horizon. In this case a naive application of the WKB approximation to the equation in this form will result in an expression that does poorly for low $\ell$, although gives the correct classical phase shift at large $\ell$. The origin of this problem as first noted by Langer~\cite{Langer:1937qr}  is that the scattering process here is defined on the line $r \ge 0$, and the behaviour of the solutions near $r=0$ is not well approximated by WKB. This problem is easily resolved performing a coordinate transformation $r = e^{\rho}$ which maps the origin at $r=0$ to $\rho=- \infty$. The correct asymptotic solution near $\rho= -\infty$ is now the exponentially decaying WKB solution. To proceed we change variables
\be
{\cal Z}_t\omega^2 \phi_\ell +e^{-(D-1) \rho} \frac{\partial}{\partial \rho}\left(  e^{(D-3)\rho} {\cal Z}_r \frac{\partial}{\partial \rho} \phi_\ell \right) - {\cal Z}_{\Omega} \ell(\ell+D-3) e^{-2 \rho}\phi_\ell = 0 \, ,\ee
and then define $\phi_\ell = e^{-(D-3)\rho/2} ({\cal Z}_r)^{-1/2} \chi_\ell$ which puts the equation in the canonical form
\be
\frac{\partial^2 \chi_\ell}{\partial \rho^2 }  = - W_\ell(\rho) \chi_\ell(\rho)\,,
\ee
where
\be
W_\ell(\rho) = \frac{1}{{\cal Z}_r}e^{2 \rho}\left[ {\cal Z}_t\omega^2-  {\cal Z}_{\Omega} \ell(\ell+D-3) e^{-2 \rho} \right]- \frac{(D-3)^2}{4} + \frac{1}{4} \frac{\( \frac{d {\cal Z}_r}{d \rho} \)^2}{({\cal Z}_r)^{2}}- \frac{1}{2} \frac{(D-3)\frac{d {\cal Z}_r }{d \rho} +\frac{d^2 {\cal Z}_r}{d \rho^2}  }{{\cal Z}_r}\nn\,.
\ee
In the usual case for which $D=4$ and ${\cal Z}_r={\cal Z}_t={\cal Z}_{\Omega}=1$ this gives
\ba
W_\ell^{D=4, \Z=1}=e^{2\rho}\omega^2-\ell (\ell+1)-\frac14 =e^{2\rho}\omega^2-\(\ell+\frac 12\)^2\,.
\ea
As pointed out by Langer, this corresponds to using the standard WKB formula with the replacement $\ell(\ell+1) \rightarrow (\ell+1/2)^2$ which is relevant at low $\ell$.  \\

The turning point, i.e. what is interpreted classically as the point of closest approach, is defined by $W_\ell(\rho_t)=0$. For $\rho < \rho_t$ we have $W_\ell <0$ and the desired WKB solution is that one that decays exponentially as $\rho \rightarrow -\infty$
\be\label{decayingsolution}
\chi_\ell \approx  \frac{\bar \chi}{(-W_\ell)^{1/4}}e^{- \int_{\rho}^{\rho_t} \ \sqrt{-W_\ell} \, \d \rho}\,,
\ee
for some normalization constant $\bar \chi$.
Using the WKB matching formula this matches onto for $\rho > \rho_t$,
\ba
\chi_\ell &\approx&  \frac{\bar \chi}{(W_\ell)^{1/4}} \sin \( \int_{\rho_t}^{\rho}  \sqrt{W_\ell} \, \d \rho + \frac{\pi}{4} \) \\
&\approx&  \frac{\bar \chi}{(W_\ell)^{1/4}} \sin \( \int_{r_t}^{r}  \sqrt{\frac{1}{{\cal Z}_r}\left( {\cal Z}_t\omega^2  -  {\cal Z}_{\Omega} \frac{(\ell+(D-3)/2)^2}{r^2} -\frac{1}{4 r^2} {\cal Z}_r\beta_R(r) \right)} \, \d r + \frac{\pi}{4} \)\,, \nn
\ea
where $r_t$ is the turning point expressed in terms of $r$ and
\be
\beta_R(r) =  (D-3)^2\(1-\frac{{\cal Z}_{\Omega} }{{\cal Z}_r}\)- r^2 \frac{\( \frac{\p {\cal Z}_r}{\p r} \)^2}{({\cal Z}_r)^2}+ 2 \frac{\((D-2)r \frac{\p {\cal Z}_r }{\p r} +r^2 \frac{\p^2 {\cal Z}_r}{\p r^2} \) }{{\cal Z}_r} \, .
\ee

In the idealized case in which all components $Z_I$ asymptote to unity faster than $1/r$, then the scattering phase shifts are determined by requiring that this solution has the asymptotic form
\be\label{asymp1}
\chi_\ell \propto  \( e^{2i \delta_\ell } e^{i \omega r} + e^{i \pi (D-2)/2} e^{i \pi \ell} e^{-i \omega r}\) \, .
\ee
Performing the comparison we obtain the standard WKB formula for the partial wave phase shifts
\ba\label{phaseshift1}
\delta_\ell(\omega) &=& \int_{r_t}^{\infty} \d r  \left( \sqrt{\frac{1}{{\cal Z}_r}\left( {\cal Z}_t\omega^2  -  {\cal Z}_{\Omega} \frac{(\ell+(D-3)/2)^2}{r^2} -\frac{1}{4 r^2} {\cal Z}_r\beta_R(r) \right)} - \omega \right) \\
&-& \omega r_t + \frac{\pi}{2} (\ell+(D-3)/2)\nn\,.
\ea
The total Eisenbud-Wigner time delay for each partial wave is then given by
\ba
\Delta T_\ell & = &2 \frac{\p \delta_\ell(\omega)}{\p \omega} \\
&=&2  \int_{r_t}^{\infty} \d r  \left( \frac{{\cal Z}_t \omega}{\sqrt{{\cal Z}_r\left( {\cal Z}_t\omega^2 -  {\cal Z}_{\Omega} \frac{(\ell+(D-3)/2)^2}{r^2} -\frac{1}{4 r^2 } {\cal Z}_r\beta_R(r) \right)}} - 1 \right) - 2 r_t  \, .
\label{TD10}
\ea
This simplifies in the limit of large $\ell$, fixed apparent impact parameter $b=(\ell+(D-3)/2)/\omega$, so that $\beta_R(r)$ may be neglected we find
\ba\label{TD1}
\Delta T_\ell  &=& 2 \int_{r_t}^{\infty}  \left( \frac{{\cal Z}_t }{\sqrt{{\cal Z}_{r}\left({\cal Z}_t -  {\cal Z}_{\Omega } \frac{b^2}{r^2}  \right)}} - 1 \right)  \d r -  2 r_t  \, , \\
&=&2 \int_{r_t}^{\infty}   \left( \frac{{\cal Z}_t }{\sqrt{{\cal Z}_{r}\left({\cal Z}_t -  {\cal Z}_{\Omega } \frac{b^2}{r^2}  \right)}} \right)\d r-2 \int_{b}^{\infty} \frac{1 }{\sqrt{1 -  \frac{b^2}{r^2} }}\d r  \,,
\ea
which is the `classical' time-delay result for a particle moving along a null geodesic in the metric $Z_{\mu\nu}$. \\

\subsection{Dealing with a Horizon}
\label{sec:horizon}
To deal with spacetimes with a horizon, it is necessary to modify the Langer transformation and instead of taking $r=0$ to $-\infty$, we map the horizon to $-\infty$. This is achieved by defining an analogue of the tortoise coordinates for which the two dimensional $r,t$ metric is conformally flat, i.e. for which
\be
Z_t^{-1/2} Z_r^{-1/2} \d r  = \d \hat r \, ,
\ee
so that the metric takes the form $\d s^2 =Z_t (-\d t^2+ \d \hat r^2) +  Z_{\Omega} r^2 \d \Omega^2$.
The integration constant in this change of variables can be fixed at infinity for $D>4$ as
\be
\hat r = r - \int^{\infty}_r \d r \(Z_t^{-1/2} Z_r^{-1/2}-1\) \, .
\ee
For $D=4$ there is a logarithmic divergence in this integral and an arbitrary finite comparison scale must be chosen \ref{Shapiro4d}.
Then with the choice $\phi_\ell = Z_{\Omega}^{-(D-2)/4}r^{-(D-2)/2}\chi_\ell $, the wave equation becomes
\be
\frac{\d^2 \chi_\ell}{\d \hat r^2}  + \omega^2\chi_\ell -V_{\rm eff}(\hat r) \chi_\ell=0 \, ,
\ee
with the effective potential
\be\label{effectivepotential}
V_{\rm eff}(\hat r) =\frac{Z_t}{Z_{\Omega} r^2} \ell(\ell+(D-3)) + \frac{1}{\gamma} \frac{\d^2 \gamma}{\d \hat r^2} \, ,
\ee
with $\gamma = Z_{\Omega}^{(D-2)/4}r^{(D-2)/2}$. In the familiar case of the $D=4$ Schwarzschild solution this is
\be\label{potentialtortoise}
V_{\rm eff}(\hat r) = \(1-\frac{r_g}{r}\) \(\frac{\ell(\ell+1)}{r^2} +\frac{r_g}{r^3}\) \, .
\ee
The asymptotic form of the mode functions is in these new coordinates conventional
\be\label{asymp2}
\chi_\ell \propto  \( e^{2i \hat \delta_\ell } e^{i \omega \hat r} - (-1)^\ell e^{-i \hat \omega r}\) \, ,
\ee
which gives the phase shift
\be\label{phaseshiftrhat}
\hat \delta_\ell(\omega) =  \int_{\hat r_t}^{\infty} \d \hat r  \(  \sqrt{\left( \omega^2-  V_{\rm eff}(\hat r) \) }-\omega \) - \omega \hat r_t + \frac{\pi}{2}(\ell+(D-3)/2) \, .
\ee
In deriving this we assume that $r_t$ is sufficiently larger than the peak of the potential $V_{\rm eff}(\hat r)$, which for example in $D=4$ occurs at approximately $r=3r_g/2$ for large $\ell$, so that there is sufficient barrier that \eqref{decayingsolution} is approximately the solution inside the potential barrier. Clearly as the turning point approaches the top of the barrier we must take appropriate consideration of the absorbed waves \cite{PhysRevD.35.3621,Sanchez:1976fcl}. Translated back into the original coordinates this is
\ba\label{phaseshift3}
\hat \delta_\ell(\omega) &=&  \int_{r_t}^{\infty} \d \hat r  \(  \sqrt{\frac{1}{{\cal Z}_r}\left( {\cal Z}_t\omega^2 -  {\cal Z}_{\Omega} \frac{\ell(\ell+(D-3))}{r^2} -\frac{1}{4 r^2 }{\cal Z}_r \beta_H(r) \right)}-\omega \) \nn\\
&-& \omega r_t + \frac{\pi}{2}(\ell+(D-3)/2) \, ,
\ea
which gives a time delay
\be
\Delta T_\ell  =2  \int_{r_t}^{\infty} \d r  \left( \frac{{\cal Z}_t \omega}{\sqrt{{\cal Z}_r\left( {\cal Z}_t\omega^2 -  {\cal Z}_{\Omega} \frac{\ell(\ell+(D-3))}{r^2} -\frac{1}{4 r^2 } {\cal Z}_r \beta_H(r) \right)}} - 1 \right) - 2 r_t  \, ,
\ee
where now
\be
\beta_H(r) =\frac{4 r^2}{Z_r Z_t} \frac{1}{\gamma} \frac{\d^2 \gamma}{\d \hat r^2} \, .
\ee
This expression is to be compared with \eqref{TD10}. The two differ only in the sub-leading semiclassical contribution which accounts for the different boundary conditions describing the two different physical situations.

\section{Eikonal as a limit of Semiclassical Phase Shifts}\label{app:eikonal}

Since many discussions of causality in effective field theories are phrased in the eikonal or shockwave (Penrose limit) approximation (see for example Refs.~\cite{Camanho:2014apa,Horowitz:1999gf,AccettulliHuber:2020oou}), it is worth showing here that the eikonal approximation can be obtained straightforwardly as a limiting case of the semiclassical approximation, and hence the latter may be regarded as more general. We begin with the phase shift relevant to the 2D conformally flat coordinates $\hat r$ \eqref{phaseshiftrhat}. We split the potential in the form of the usual centrifugal potential plus corrections
\be
V_{\rm eff}(\hat r) = \frac{b^2 \omega^2}{\hat r^2} + U_{\rm eff}(\hat r) \, ,
\ee
with $b=(\ell+(D-3)/2) \omega^{-1}$. The eikonal approximation corresponds to assuming a high energy limit $\omega^2 \gg  |U_{\rm eff}(\hat r) |$ so that we may treat $ U_{\rm eff}(\hat r) $ perturbatively. In the present relativistic context this limit is more subtle than it is in non-relativistic quantum mechanics since $U_{\rm eff}(\hat r) $ itself scales with $\omega^2$. Nevertheless its different radial dependence ensures there is always a regime in which we may imagine $\omega^2 \gg U_{\rm eff}(\hat r) $. Naively we can just perturb the square root in \eqref{phaseshiftrhat}, however this becomes problematic since the point of closest approach $\hat r_t$ is itself dependent on the potential $U_{\rm eff}(\hat r) $ and a naive expansion will lead to an ill-defined expression. The solution is to use the relation between $b$ and $\hat r_t$
\be\label{brt}
 \frac{b^2 \omega^2}{\hat r_t^2} + U_{\rm eff}(\hat r_t) =\omega^2 \, ,
\ee
to rewrite \eqref{phaseshiftrhat} in the form
\be\label{phaseshiftrhat2}
\hat \delta_\ell(\omega) =  \int_{\hat r_t}^{\infty} \d \hat r  \(  \sqrt{\left( \omega^2-  \frac{\hat r_t^2 \omega^2}{\hat r^2} - \(U_{\rm eff}(\hat r) -U_{\rm eff}(\hat r_t) \frac{\hat r_t^2}{\hat r^2}  \) \right) }-\omega \)  - \omega \hat r_t + \frac{\pi}{2}(\ell+(D-3)/2) \, ,
\ee
which admits a well defined expansion to any order, for which the first order term in $U_{\rm eff}(\hat r)$ is
\ba
\hat \delta_\ell(\omega) &=&  \int_{\hat r_t}^{\infty} \d \hat r  \(  \sqrt{\left( \omega^2-  \frac{\hat r_t^2 \omega^2}{\hat r^2} \right)} - \omega \)- \omega \hat r_t + \frac{\pi}{2}(\ell+(D-3)/2) \nn \\
& - & \frac{1}{2}\int_{\hat r_t}^{\infty} \d \hat r  \frac{1}{\sqrt{ \omega^2-  \frac{\hat r_t^2 \omega^2}{\hat r^2} }} \(U_{\rm eff}(\hat r) -U_{\rm eff}(\hat r_t) \frac{\hat r_t^2}{\hat r^2}  \) + \dots   \\
&=& \frac{\pi}{2}\omega (b-\hat r_t) -  \frac{1}{2}\int_{\hat r_t}^{\infty} \d \hat r  \frac{1}{\sqrt{ \omega^2-  \frac{\hat r_t^2 \omega^2}{\hat r^2} }} \(U_{\rm eff}(\hat r) -U_{\rm eff}(\hat r_t) \frac{\hat r_t^2}{\hat r^2}\) \, .
\label{phaseshiftrhat3}
\ea
Now from \eqref{brt}
\be
b-\hat r_t = \frac{\hat r_t^2 \omega^{-2}}{b+ \hat r_t} U_{\rm eff}(\hat r_t) \approx \frac{\hat r_t}{2 \omega^2}U_{\rm eff}(\hat r_t)  \, ,
\ee
and so substituting in \eqref{phaseshiftrhat3} we have
\be
\hat \delta_\ell(\omega)=-  \frac{1}{2 \omega }\int_{\hat r_t}^{\infty} \d \hat r  \frac{1}{\sqrt{1-  \frac{\hat r_t^2 }{\hat r^2} }} U_{\rm eff}(\hat r) + \dots \, .
\ee
To put this in a recognizable form we define $\hat r^2 = \hat r_t^2 + z^2$ and notice that to this order we may replace $\hat r_t$ with the apparent impact parameter $b$ to give
\be\label{phaseshiftrhat4}
\hat \delta_\ell(\omega)\approx \hat \delta_\ell^{\rm eik}(\omega)=-  \frac{1}{4 \omega }\int_{-\infty}^{\infty} \d z   \, U_{\rm eff}(\sqrt{b^2+z^2})  \, ,
\ee
which is clearly the eikonal result. This same expression may be derived in a more covariant manner by recognizing the in the classical limit the phase shift $\delta $ is one half the difference between the action of a relativistic particle in the curved spacetime $Z_{\mu\nu}$ and in Minkowski spacetime with the same asymptotic momenta. Writing the action for a massless relativistic particle in phase space form
\be
S = \int_{-\infty}^{\infty} \d \tau  \( p_{\mu} \frac{\d x^{\mu}}{\d \tau} - \frac{1}{2} Z^{\mu\nu} p_{\mu} p_{\nu} \)\,,
\ee
then perturbing around Minkowski we have
\ba
\hat \delta^{\rm eik} &\approx &  \frac{1}{4} \int_{-\infty}^{\infty} \d \tau \(  \delta Z_{\mu\nu} p^{\mu} p^{\nu}\)=\frac{1}{2} \int_{b}^{\infty} \d r \frac{\d \tau}{\d r} \(  \delta Z_{\mu\nu} p^{\mu} p^{\nu}\)  \, ,
\ea
where the momentum $p^{\mu}$ and velocity are their solutions in Minkowski spacetime written in radial coordinates, i.e. $\frac{d r}{\d \tau} = \omega \sqrt{1-b^2/r^2}$, $p^{t} =\omega$, $p^r = \omega \sqrt{1-b^2/r^2}$, $p^{\theta} = \omega b/r^2$ which for a spherically symmetric background in 2D conformally flat coordinates
\be
\hat \delta^{\rm eik} =-\frac{1}{4 \omega } \int_{-\infty}^{\infty} \d z \frac{\omega^2 b^2}{b^2+z^2}  \(  \delta Z_{t}(\sqrt{b^2+z^2})-\delta Z_{\Omega}(\sqrt{b^2+z^2})     \)  \, ,
\ee
which is consistent with \eqref{phaseshiftrhat4} given $U_{\rm eff}(\hat r) =\frac{b^2 \omega^2}{\hat r^2}(  \delta Z_{t}(\hat r)  -\delta Z_{\Omega}(\hat r)  )$ for large $\ell$ from \eqref{effectivepotential}.

\subsection{Shapiro time delay in $D>4$}

To illustrate the applicability of the above formula let us compute the $D$ dimensional Shapiro delay. We begin with the Schwarzschild-Tangherlini metric in $D$ dimensions
\be
\d s^2=- f(r)\, \d t^2+f(r)^{-1}\d r^2+   r^2\, \d \Omega_{D-2}^2\,,
\ee
with $f(r) = 1- \mu/r^{D-3}$ and $\mu$ related to the physical source mass $M_*$ by
\ba
\mu = \frac{16 \pi G M_*}{(D-2) \Omega_{D-2}}\,,
\ea
with $\Omega_{D-2}=2 \pi^{(D-1)/2}/\Gamma[(D-1)/2]$.
We define the 2D conformal (tortoise) coordinate $\hat r$ via
\be\label{2Dcoordinate}
\hat r = r -\int^{\infty}_r \d r \(  \frac{1}{f(r)}-1\) = r - \frac{1}{(D-4)} \frac{\mu}{r^{D-4}} + \dots
\ee
hence to leading order in $\mu$ the metric is
\be
\d s^2 \approx \(1 - \frac{\mu}{\hat r^{D-3}} \)( -\d t^2 + \d \hat r^2 )  + \( 1+ \frac{2}{(D-4)} \frac{\mu}{\hat r^{D-2}}\)\hat r^2 \d^2 \Omega_{D-2} \, ,
\ee
from which we infer the effective potential
\be
U_{\rm eff}( \hat r) = -\frac{(D-2)}{(D-4)}b^2 \omega^2 \frac{\mu}{\hat r^{D-1}} \, .
\ee
Hence the leading order contribution to the phase shift from \eqref{phaseshiftrhat4} is
\ba
\delta_\ell(\omega)&=& \frac{(D-2)}{(D-4)}b^2 \omega \frac{\mu}{4} \int_{-\infty}^{\infty} \d z   \, \frac{1}{(b^2+z^2)^{(D-1)/2}} \\
&=& \frac{(D-2)}{(D-4)}b^2 \omega \frac{\mu}{4} b^{4-D} \sqrt{\pi} \frac{\Gamma\left[\frac{D-4}{2}\right] }{\Gamma\left[\frac{D-1}{2}\right] }=\frac{ GM_* \omega }{ \pi^{(D-4)/2} b^{D-4}} \Gamma\left[\frac{D-4}{2}\right] \, .
\ea
Denoting the Mandelstam invariant $s =-p_{\rm total}^2= (M_*+\omega)^2 - \omega^2 \approx 2 M_* \omega$ for large $\omega$ this is\be
\delta_\ell(\omega)=\frac{s G}{2 \pi^{(D-4)/2} b^{D-4}} \Gamma\left[\frac{D-4}{2}\right] \, .
\ee
In this Lorentz invariant form we may easily translate it into the boosted Penrose limit form considered for example in \cite{Camanho:2014apa}\footnote{Note our definitions of $\delta_\ell(\omega)$ differ by a factor of $2$.}, and so we see the shockwave calculation reproduced precisely the leading term in the semiclassical expansion. Given the phase shift we may now define the time delay for fixed impact parameter $b$ as
\be\label{Shapiro1}
\Delta T_b = 2 \frac{\partial \delta_\ell(\omega)}{\delta \omega} \Big|_b =  \frac{2M_* G}{ \pi^{(D-4)/2} b^{D-4}} \Gamma\left[\frac{D-4}{2}\right] \, ,
\ee
which is the conventional classical Shapiro delay. However, quantum mechanically what is more meaningful is the delay for fixed $\ell$, which is the usual definition of the Eisenbud-Wigner time delay for partial waves
\be\label{Shapiro2}
\Delta T_\ell = 2 \frac{\partial \delta_\ell(\omega)}{\delta \omega} \Big|_\ell =  \frac{2 M_* G(D-3)}{ \pi^{(D-4)/2} b^{D-4}} \Gamma\left[\frac{D-4}{2}\right] \, .
\ee
Both expressions are finite and positive and the former plays a crucial role in the discussion in \cite{Camanho:2014apa}. \\

An alternative derivation of this result is to consider
\be
\Delta T_\ell =2 \int_{r_t}^{\infty}    \frac{\d r }{f(r) \sqrt{1 -  f(r) \frac{b^2}{r^2}  }} -2 \int_{b}^{\infty}    \frac{\d r}{\sqrt{1 -  \frac{b^2}{r^2}  }} \, ,
\ee
and following the approach described in appendix \ref{app:Timedelaycalc} to leading order in an expansion in $G$ this is
\ba \label{HigherD}
\Delta T_\ell & \approx& 2 \int_{b}^{\infty} \d r   \frac{1 }{\sqrt{1 -  \frac{b^2}{r^2}  }} \frac{\mu}{2 b^2 }\frac{1}{r^{D-3}}\left[3 b^2 (D-4) -2 r^2(D-6) \right] \\
&=& \frac{b^{4-D} (D-2) \mu \sqrt{\pi} \, \Gamma[\frac{D-4}{2}]}{2 \Gamma[\frac{D-3}{2}]}  =\frac{2GM_* (D-3)}{\pi^{(D-4)/2} b^{D-4}} \Gamma\left[\frac{D-4}{2}\right]  \, ,
\ea
which agrees with \eqref{Shapiro2}. Strictly speaking the integral \eqref{HigherD} is only convergent for $D>6$ but the final result is finite and correct for $D>4$.

\subsection{Shapiro time delay in $D=4$}\label{Shapiro4d}

The result \eqref{TD1} only applies if the effective metric asymptotes to Minkowski faster than $1/r$ to ensure convergence of the integral. In four dimensional GR this is never the case and \eqref{TD1} is divergent, a result of the massless nature of the graviton.  This dimensionality dependence may be related to that of  Schwarzschild causality established in \cite{Cameron:2020itp}.
The resolution is to account for the attractive Coulomb distortion of the asymptotic wavefunctions by modifying \eqref{asymp1} to
\be\label{asymp2}
\chi_\ell \propto  \( e^{2i \delta'_\ell } e^{i \omega r+ i \omega \alpha \ln(2 \omega r)} - (-1)^\ell e^{-i \omega r-i \omega \alpha \ln(2 \omega r)}\) \, ,
\ee
which replaces \eqref{phaseshift1} with the finite phase shift
\ba\label{phaseshift2}
\delta'_\ell(\omega) &=& \lim_{R \rightarrow \infty} \Bigg[ \int_{r_t}^{R} \d r   \sqrt{\frac{1}{{\cal Z}_r}\left( {\cal Z}_t\omega^2  -  {\cal Z}_{\Omega} \frac{(\ell+1/2)^2}{r^2} -\frac{1}{4 r^2} {\cal Z}_r\beta_R(r) \right)}   \\
&+& \frac{\pi}{2} (\ell+1/2)- \omega R-  \omega \alpha \ln(2 \omega R) \Bigg] \, ,\nn
\ea
for regular geometries. For geometries with a horizon, the analogue of  \eqref{phaseshift3} is replaced by \cite{Sanchez:1976fcl}
\ba\label{phaseshift20}
\hat \delta_\ell(\omega) &=&   \lim_{R \rightarrow \infty} \Bigg[ \int_{r_t}^{R} \d \hat r  \(  \sqrt{\frac{1}{{\cal Z}_r}\left( {\cal Z}_t\omega^2 -  {\cal Z}_{\Omega} \frac{\ell(\ell+1)}{r^2} -\frac{1}{4 r^2 }{\cal Z}_r \beta_H(r) \right)}-\omega \) \\
&-& \omega r_t + \frac{\pi}{2}\(\ell+\frac{1}{2}\) -  \omega \alpha \ln(2 \omega R) \Bigg]\, ,\nn
\ea
where the Coulomb distortion scale $\alpha$ is defined by
\be
\alpha =\lim_{r \rightarrow \infty}
\left[ \frac{r}{2} \left( \frac{{\cal Z}_t}{{\cal Z}_r} -1 \right) \right] \,,
\ee
which in the standard case of an asymptotically Schwarzschild geometry is the Schwarzschild radius $\alpha = r_g$. In particular evaluating \eqref{phaseshift20} on Schwarzschild to leading order in an expansion in $r_g/b$ gives  \cite{Sanchez:1976fcl},
\be
\hat \delta_\ell(\omega) = - \omega r_g \ln(\ell+1/2)-\frac{1}{2} \omega r_g \(1+\frac{1}{(\ell+1/2)^2}\)+\dots \, .
\ee
Unlike the result in dimensions $D>4$ \eqref{HigherD}, this corresponds to a time advance
\be
\Delta T_\ell = - r_g \(  2 \ln(\ell+1/2)+ \(1+\frac{1}{(\ell+1/2)^2}\)\) \, .
\ee
We should not interpret this as any violation of causality though. Firstly, even in non relativistic quantum mechanics time advances do occur, as is well known in the example of scattering from a hard sphere where, for a sphere of radius $a$, the scattered wave of speed $v$ is reflected at a time $a/v$ before a free wave would reach the center at $r=0$ and so is reduced in travel time by $2a/v$. For this reason, Wigner's original causality condition for $S$-wave scattering is stated as \cite{Wigner:1955zz,DECARVALHO200283}
\be
\Delta T_{\ell =0} \ge - 2 a/v \, ,
\ee
up to fluctuations at the scale $\omega^{-1}$. The implication is that unlike in higher dimensions, the black hole acts like scattering off of a hard sphere with an $\ell$ dependent radius of order $r_g$. However this is also at the scale of the ambiguity in the definition of the phase due to the divergence from the Coulombic behaviour, i.e. the logarithmic term included in \eqref{phaseshift20} and so we should be careful to read too much into this.
Indeed in defining the tortoise coordinates analogous to \eqref{2Dcoordinate} we are faced with a logarithmic divergence which must be cutoff at a scale $C$
\be
\hat r= r - \int^C_{r} \d r \frac{1}{1-\frac{r_g}{r}} = r - r_g \ln((r-r_g)/(C-r_g))  \, .
\ee
In pure Schwarzschild, $C$ is usually fixed to be $2r_g$ but in a spacetime which is only asymptotically Schwarzschild there is no requirement for this particular choice. The inherent logarithmic ambiguity in the $\hat r$ coordinate translates into the same ambiguity in the phase shift and hence time delay. As such we will content ourselves with determining time delays relative to the asymptotic Schwarzschild for which the ambiguity related to the definition of the phase shift cancels out. This corresponds to the classical criterion \cite{Gao:2000ga}.
It is noteworthy that these issues are entirely avoided in higher dimensions.
Another way to understand these subtleties is to introduce an IR regulator through an effective mass $\mu$ by replacing the Coulombic $r_g/r$ with a Yukawa form $r_g e^{-\mu r}/r$. This renders the naive phase shift definition \eqref{phaseshift1} and \eqref{phaseshift3} finite, and gives a time delay logarithmically sensitive to the IR cutoff. This divergence will cancel in considering time delay differences as we do throughout, allowing us to take the limit $\mu \rightarrow 0$.
\\

Alternatively, from a classical point of view we can regulate the divergence by asking for the time delay for a trajectory that begins and ends at finite radii $r_b$ and $r_e$ respectively relative to the same result in Minkowski spacetime with impact parameter $b$
\ba\label{TD2}
\Delta T_\ell  &=&  \int_{r_t}^{r_e} \d r  \frac{{\cal Z}_t }{\sqrt{{\cal Z}_{r}\left({\cal Z}_t -  {\cal Z}_{\Omega } \frac{b^2}{r^2}  \right)}}  - \int_{r_t}^{r_e} \d r   \frac{1}{\sqrt{1-\frac{b^2}{r^2}}} \\
&+& \int_{r_t}^{r_b} \d r  \frac{{\cal Z}_t }{\sqrt{{\cal Z}_{r}\left({\cal Z}_t -  {\cal Z}_{\Omega } \frac{b^2}{r^2}  \right)}}  - \int_{r_t}^{r_b} \d r   \frac{1}{\sqrt{1-\frac{b^2}{r^2}}}  \, .\nn
\ea
Evaluating this for a Schwarzschild geometry we have the well known Shapiro time-delay written in terms of coordinate time $t$ \cite{Shapiro:1964uw}
\be
\Delta T_\ell^g = \int_{r_t}^{r_e} \d r \(\frac{1}{f(r)\sqrt{1-\frac{b^2}{r^2}f(r) }}\)+ \int_{r_t}^{r_b}  \d r \(\frac{1}{f(r)\sqrt{1-\frac{b^2}{r^2}f(r) }}\)-(r_e^2-b^2)^{1/2} - (r_b^2-b^2)^{1/2}   \, ,
\ee
with $f(r)= 1- r_g/r$ which to first order in $r_g/r_t$ is\footnote{This result is more often quoted as the total time signalling time back and forth between $r_b$ and $r_e$ in the proper time of an observer at $r_b$ which in our notation is $\Delta \tau_{b \rightarrow e \rightarrow b}=2 (1- r_g/r_b) ((r_e^2-b^2)^{1/2} + (r_b^2-b^2)^{1/2} +\Delta T_\ell^g )$. }
\ba
\Delta T_\ell^g &=& r_g \ln\(\frac{r_e+\sqrt{r_e^2-r_t^2}}{r_t} \)  +r_g \ln\(\frac{r_b+\sqrt{r_b^2-r_t^2}}{r_t} \)  +r_g \sqrt{\frac{r_e-r_t}{r_e+r_t}}+r_g \sqrt{\frac{r_b-r_t}{r_b+r_t}}\nn  \\
&&+(r_e^2-r_t^2)^{1/2} + (r_b^2-r_t^2)^{1/2} -  (r_e^2-b^2)^{1/2} - (r_b^2-b^2)^{1/2}  +  {\cal O}(r_g^2/r_t) \, .
\ea

\section{Time Delay Corrections}\label{app:Timedelaycalc}

We are generally interested in corrections to the time-delay from EFT corrections to the effective background geometry and hence consider the geometry \eqref{eq:ZZ} with now $Z_{\mu\nu}  \rightarrow Z_{\mu\nu} + \delta Z_{\mu\nu}$.
In terms of the corrected factors $\Z_I\to \Z_I+\delta \Z_I$ as defined below the wave equation \eqref{eq:wave},
the correction to the time-delay is defined by
\ba\label{TD2}
\Delta T_\ell^{\rm EFT}  &=&  2 \int_{r_t+\delta r_t}^{\infty} \d r  \left( \frac{{\cal Z}_t +\delta {\cal Z}_t }{\sqrt{({\cal Z}_{r}+ \delta {\cal Z}_{r})\left({\cal Z}_t+\delta  {\cal Z}_t- ( {\cal Z}_{\Omega }+ \delta {\cal Z}_{\Omega } ) \frac{b^2}{r^2}  \right)}} \right) \nn \\
&-&2\int_{r_t}^{\infty} \d r  \( \frac{{\cal Z}_t }{\sqrt{{\cal Z}_{r}\left({\cal Z}_t -  {\cal Z}_{\Omega } \frac{b^2}{r^2}  \right)}}  \) \, ,
\ea
where $r_t$ is the turning point in GR (where ${\cal Z}_t(r_t)  r_t^2 =  {\cal Z}_{\Omega }(r_t) b^2$) and $r_t+\delta r_t$ is the turning point associated with the effective metric $Z_{\mu\nu} + \delta Z_{\mu\nu}$.

For simplicity, we shall denote the integrant in GR as $\A$ and that in the EFT as $\A +\delta \A$,
\ba
\A(r)&=& \frac{{\cal Z}_t }{\sqrt{{\cal Z}_{r}\left({\cal Z}_t -  {\cal Z}_{\Omega } \frac{b^2}{r^2}  \right)}}\\
\delta \A(r)&=& -\frac{1}{2}\frac{\A^3}{\Z_t^3}
\(\Z_t(\Z_t\delta \Z_r-\Z_r \delta \Z_t)
-(\Z_t \Z_\Omega \delta \Z_r-2 \Z_r \Z_\Omega \delta \Z_t +\Z_r \Z_t \delta \Z_\Omega)\frac{b^2}{r^2}\)\,.
\ea
To determine the correction $\Delta T_\ell^{\rm EFT}$,  we perform in each integral a distinct coordinate transformation $r \rightarrow \rho$. For the second integral in \eqref{TD2} we perform a change of variable defined as $r=R(\rho)$ so that the relation \eqref{eq:RI2} below be satisfied and for the first integral in \eqref{TD2} we perform a change of variable $r=R(\rho)+\delta R(\rho)$ defined by
\ba
&& I_2 : 1-\frac{1}{\rho^2} = \A^{-2}(r)\Bigg|_{r=R(\rho)}\, ,
\label{eq:RI2}\\
&& I_1 : 1-\frac{1}{\rho^2} = \(\A(r)+\delta \A(r)\)^{-2}\Bigg|_{r=R(\rho)+\delta R(\rho)} \, ,
\label{eq:RI1}
\ea
in terms of which the time delay correction is
\ba
\Delta T_\ell^{\rm EFT}=
\label{Timeintegral2}
2\int_1^{\infty}  \frac{\d \rho}{\sqrt{1-1/\rho^2} }\left( \frac{\d \delta R(\rho)}{\d \rho}   \right) \, .
\ea
To first order in $\delta Z_{\mu\nu}$, the function $\delta R(\rho)$ is determined by perturbing \eqref{eq:RI1}, giving
\ba
\label{eq:deltaR}
 \delta R(\rho) =- \frac{\delta \A(r)}{\A'(r)}\Bigg|_{r=R(\rho)} \, .
\ea
Substituting this expression back into \eqref{Timeintegral2} and using \eqref{eq:RI2} to change variables back to $r$ gives
\ba
\label{TD5}
\Delta T_\ell^{\rm EFT} = -2 \int_{r_t}^{\infty} \d r \A(r)\frac{\d }{\d r}\(\frac{\delta \A(r)}{\A'(r)}\)\,.
\ea

\newpage

\bibliographystyle{JHEP}
\bibliography{references}

\end{document}